\documentclass[twocolumn]{aastex7}

\newcommand{\Yp}{Y_\mathrm{P}}
\newcommand{\yp}{y_\mathrm{P}}
\newcommand{\neff}{N_\mathrm{eff}}
\newcommand{\xie}{\xi_\mathrm{e}}
\newcommand{\err}[3]{#1^{+{#2}}_{-#3}}
\usepackage{amsmath}
\usepackage{bm} % bold math symbols
\usepackage{xcolor} % for \textcolor
\usepackage{rotating}

\newcommand{\boldtext}[1]{\ifmmode\bm{#1}\else\textbf{#1}\fi}
\begin{document}

\shorttitle{Primordial Helium Abundance}

\shortauthors{Yanagisawa et al.}

\title{
EMPRESS. XV. A New Determination of the Primordial Helium Abundance\\
Suggesting a Moderately Low $Y_\mathrm{P}$ Value
}

\author[0009-0006-6763-4245]{Hiroto Yanagisawa}
\affiliation{Institute for Cosmic Ray Research, The University of Tokyo, 5-1-5 Kashiwanoha, Kashiwa, Chiba 277-8582, Japan}
\affiliation{Department of Physics, Graduate School of Science, The University of Tokyo, 7-3-1 Hongo, Bunkyo, Tokyo 113-0033, Japan}
\email{yana@icrr.u-tokyo.ac.jp}

\author[0000-0002-1049-6658]{Masami Ouchi}
\affiliation{National Astronomical Observatory of Japan, National Institutes of Natural Sciences, 2-21-1 Osawa, Mitaka, Tokyo 181-8588, Japan}
\affiliation{Institute for Cosmic Ray Research, The University of Tokyo, 5-1-5 Kashiwanoha, Kashiwa, Chiba 277-8582, Japan}
\affiliation{Department of Astronomical Science, SOKENDAI (The Graduate University for Advanced Studies), 2-21-1 Osawa, Mitaka, Tokyo, 181-8588, Japan}
\affiliation{Kavli Institute for the Physics and Mathematics of the Universe (WPI), University of Tokyo, Kashiwa, Chiba 277-8583, Japan}
\email{ouchims@icrr.u-tokyo.ac.jp}

\author{Akinori Matsumoto}
\affiliation{Institute for Cosmic Ray Research, The University of Tokyo, 5-1-5 Kashiwanoha, Kashiwa, Chiba 277-8582, Japan}
\affiliation{Department of Physics, Graduate School of Science, The University of Tokyo, 7-3-1 Hongo, Bunkyo, Tokyo 113-0033, Japan}
\email{matsumoto-akinori489@g.ecc.u-tokyo.ac.jp}

\author[0000-0001-9991-7051]{Masahiro Kawasaki}
\affiliation{Institute for Cosmic Ray Research, The University of Tokyo, 5-1-5 Kashiwanoha, Kashiwa, Chiba 277-8582, Japan}
\email{kawasaki@icrr.u-tokyo.ac.jp}

\author[0000-0003-2879-1724]{Kai Murai}
\affiliation{Department of Physics, Tohoku University, Sendai, Miyagi 980-8578, Japan}
\email{kai.murai.e2@tohoku.ac.jp}

\author[0000-0003-2965-5070]{Kimihiko Nakajima}
\affiliation{Institute of Liberal Arts and Science, Kanazawa University, Kakuma-machi, Kanazawa, 920-1192, Ishikawa, Japan}
\affiliation{National Astronomical Observatory of Japan, 2-21-1 Osawa, Mitaka, 181-8588, Tokyo, Japan}
\email{knakajima@staff.kanazawa-u.ac.jp}

\author[0000-0003-3764-8612]{Kazunori Kohri}
\affiliation{National Astronomical Observatory of Japan, National Institutes of Natural Sciences, 2-21-1 Osawa, Mitaka, Tokyo 181-8588, Japan}
\affiliation{School of Physical Sciences, Graduate University for Advanced
Studies (SOKENDAI), 2-21-1 Osawa, Mitaka, Tokyo 181-8588, Japan}
\affiliation{Theory Center, IPNS, KEK, 1-1 Oho, Tsukuba, Ibaraki 305-0801, Japan}
\affiliation{Kavli Institute for the Physics and Mathematics of the Universe (WPI), University of Tokyo, Kashiwa, Chiba 277-8583, Japan}
\email{kazunori.kohri@nao.ac.jp}

\author[0000-0001-6958-7856]{Yuma Sugahara}
\affiliation{Waseda Research Institute for Science and Engineering, Faculty of Science and Engineering, Waseda University, 3-4-1 Okubo, Shinjuku, Tokyo 169-8555, Japan}
\affiliation{Department of Physics, School of Advanced Science and Engineering, Faculty of Science and Engineering, Waseda University, 3-4-1 Okubo, Shinjuku, Tokyo 169-8555, Japan}
\email{sugayu@aoni.waseda.jp}

\author[0000-0001-7457-8487]{Kentaro Nagamine}
\affiliation{Theoretical Astrophysics, Department of Earth and Space Science, Graduate School of Science, Osaka University, Toyonaka, Osaka 560-0043, Japan}
\affiliation{Kavli Institute for the Physics and Mathematics of the Universe (WPI), University of Tokyo, Kashiwa, Chiba 277-8583, Japan}
\affiliation{Department of Physics \& Astronomy, University of Nevada, Las Vegas, 4505 S. Maryland Pkwy, Las Vegas, NV 89154-4002, USA}
\affiliation{Theoretical Joint Research, Forefront Research Center, Graduate School of Science, Osaka University, Toyonaka, Osaka, 560-0043, Japan}
\email{kn@astro-osaka.jp}

\author[0000-0002-4937-4738]{Ichi Tanaka}
\affiliation{Subaru Telescope, National Astronomical Observatory of Japan, 650 North A’ohoku Place, Hilo, Hawaii, 96720, USA}
\email{ichi@naoj.org}

\author[0000-0002-1418-3309]{Ji Hoon Kim}
\affiliation{SNU Astronomy Research Center, Department of Physics \& Astronomy, Seoul National University, 1 Gwanak-ro, Gwanak-gu, \\Seoul 08826, Republic of Korea}
\email{jhkim.astrosnu@gmail.com}

\author[0000-0001-9011-7605]{Yoshiaki Ono}
\affiliation{Institute for Cosmic Ray Research, The University of Tokyo, 5-1-5 Kashiwanoha, Kashiwa, Chiba 277-8582, Japan}
\email{ono@icrr.u-tokyo.ac.jp}

\author[0009-0000-1999-5472]{Minami Nakane}
\affiliation{Institute for Cosmic Ray Research, The University of Tokyo, 5-1-5 Kashiwanoha, Kashiwa, Chiba 277-8582, Japan}
\affiliation{Department of Physics, Graduate School of Science, The University of Tokyo, 7-3-1 Hongo, Bunkyo, Tokyo 113-0033, Japan}
\email{nakanem@icrr.u-tokyo.ac.jp}

\author[0000-0002-5045-6052]{Keita Fukushima}
\affiliation{Department of Physics \& Astronomy, Seoul National University, Seoul 08826, Republic of Korea}
\email{fukushima@snu.ac.kr}

\author[0000-0002-6047-430X]{Yuichi Harikane}
\affiliation{Institute for Cosmic Ray Research, The University of Tokyo, 5-1-5 Kashiwanoha, Kashiwa, Chiba 277-8582, Japan}
\email{hari@icrr.u-tokyo.ac.jp}

\author[0000-0002-5661-033X]{Yutaka Hirai}
\affiliation{Department of Community Service and Science, Tohoku University of Community Service and Science, 3-5-1 Iimoriyama, Sakata, Yamagata 998-8580, Japan}
\email{yutaka.hirai@koeki-u.ac.jp}

\author[0000-0001-7730-8634]{Yuki Isobe}
\affiliation{Kavli Institute for Cosmology, University of Cambridge, Madingley Road, Cambridge, CB3 0HA, UK}
\affiliation{Cavendish Laboratory, University of Cambridge, 19 JJ Thomson Avenue, Cambridge, CB3 0HE, UK}
\affiliation{Waseda Research Institute for Science and Engineering, Faculty of Science and Engineering, Waseda University, 3-4-1, Okubo, Shinjuku, Tokyo 169-8555, Japan}
\email{yi264@cam.ac.uk}

\author[0000-0002-3801-434X]{Haruka Kusakabe}
\affiliation{Department of General Systems Studies, Graduate School of Arts and Sciences, The University of Tokyo, 3-5-1 Komaba, Meguro, Tokyo, 153-0041, Japan}
\email{haruka.kusakabe.takeishi@gmail.com}

\author[0000-0003-3228-7264]{Masato Onodera}
\affiliation{Subaru Telescope, National Astronomical Observatory of Japan, National Institutes of Natural Sciences (NINS), 650 North Aohoku Place, Hilo, HI 96720, USA}
\affiliation{Department of Astronomical Science, SOKENDAI (The Graduate University for Advanced Studies), Osawa 2-21-1, Mitaka, Tokyo, 181-8588, Japan}
\email{monodera@naoj.org}

\author[0000-0002-1690-3488]{Michael Rauch}
\affiliation{The Observatories of the Carnegie Institution for Science, 813 Santa Barbara Street, Pasadena, CA 91101, USA}
\email{mr@carnegiescience.edu}

\author[0000-0002-1319-3433]{Hidenobu Yajima}
\affiliation{Center for Computational Sciences, University of Tsukuba, Ten-nodai, 1-1-1 Tsukuba, Ibaraki 305-8577, Japan}
\email{yajima@ccs.tsukuba.ac.jp}

\begin{abstract}
We present a new constraint on the primordial helium abundance, $Y_\mathrm{P}$, based on Subaru observations. A major source of uncertainty in previous $Y_\mathrm{P}$ determinations is the lack of extremely metal-poor galaxies (EMPGs; $0.01-0.1\,Z_\odot$), which have metallicities a few to ten times lower than the metal-poor galaxies (MPGs; $0.1-0.4\,Z_\odot$) predominantly used in earlier studies, requiring substantial extrapolation to zero metallicity. Here, we perform Subaru near-infrared spectroscopy of 29 galaxies, including 14 EMPGs. By incorporating existing optical spectra, we derive He/H for each galaxy using photoionization modeling of helium and hydrogen emission lines, including the He \textsc{i} 10830\AA \, line to break the density--temperature degeneracy. After carefully selecting galaxies with robust He/H determinations, and adding 58 galaxies from previous studies, we obtain $Y_\mathrm{P} = 0.2402^{+0.0040}_{-0.0040}$. This $Y_\mathrm{P}$ value is $\sim1\sigma$ lower than most of the previous estimates, but agrees with recent determinations using EMPGs and the CMB constraint from the Atacama Cosmology Telescope (ACT) experiment. Our result indicates $N_\mathrm{eff} = 2.54^{+0.20}_{-0.25}$, showing a mild ($\sim2\sigma$) tension with the Standard Model and Planck results. These tensions may suggest a nonzero lepton asymmetry $(\xi_\mathrm{e}\neq0)$, which would alleviate the tension with $\xi_\mathrm{e} = 0.05^{+0.02}_{-0.03}$. More observations of EMPGs and further assessments of systematic uncertainties are essential to test the potential tension more rigorously.

\end{abstract}

\keywords{}

\section{Introduction}
The $\Lambda$CDM model has successfully explained the observational results of the cosmic microwave background (CMB), the large-scale structures, and the expansion of the universe.
However, there is a $\sim 5\sigma$ tension between the Hubble parameter $H_0$ obtained from the CMB \citep{Planck2020} and local probes such as type Ia supernovae (SNe Ia) data, which is the so-called Hubble tension \citep{Riess_2022}. 
The Hubble tension suggest physics beyond the $\Lambda$CDM model. For example, if the effective number of neutrino species $\neff$ is larger than the standard value of 3.044 \citep{Froustey+2020, Bennett+2021}, the Hubble tension can be mitigated because the large $\neff$ increases the expansion rate of the early universe  \citep{Vagnozzi_2020, Seto&Toda_21}. The larger value of $\neff$ is realized by the extra radiation component such as sterile neutrino \citep{DiValentino+2013}.

The $\neff$ value can be constrained from the abundances of the primordial $^{4}$He (hereafter He) produced by the Big Bang nucleosynthesis (BBN; \citealt{Cyburt+2016}). A large $\neff$ value corresponds to a high expansion rate of the early universe, which increases the neutron-to-proton abundance ratio. Because the neutrons form into the He atoms, the primordial He abundance strongly depend on $\neff$. 

The primordial He abundance in mass fraction, $\Yp$, 
is evaluated by observations of metal-poor galaxies, whose chemical compositions are nearly primordial
(\citealp{Izotov+2014, Aver_2015, Peimbert+2016, Velardi+2019, Fernandez+2019, Hsyu+2020, Kurichin+2021, Matsumoto+2022, Skillman+2026, Rogers+2026, Weller+2026, Aver+2026, Yeh+2026}). Especially, extremely metal-poor galaxies (EMPGs), which have a metallicity smaller than 10\% $Z_\odot$, are important for the precise $\Yp$ measurement. To determine a helium abundance in these galaxies, it is necessary to model physical parameters in the galaxies, which include
%such as 
the electron temperature and density. While there is a degeneracy between the electron temperature and density, \cite{Izotov+2014} and \cite{Aver_2015} demonstrate the importance of the near-infrared 
He \textsc{i} $\lambda$ 10830 \AA\ emission 
line to break this degeneracy due to its sensitivity on the electron density. \cite{Hsyu+2020} use 54 metal-poor galaxies, 8 of which have the He {\sc i} $\lambda$ 10830 \AA\ emission line, and obtain $\Yp = 0.2436^{+0.0039}_{-0.0040}$. \cite{Hsyu+2020} combine their result with the primordial deuterium abundance measurement of $\mathrm{(D/H)_P} = (2.527 \pm 0.030) \times 10^{-5}$ \citep{Cooke+2018_Dp}, and obtain $\neff = 2.85^{+0.28}_{-0.25}$, which is consistent with the standard value of $\neff = 3.044$ within the $1\sigma$ level.

Recently, \cite{Matsumoto+2022} add 5 EMPGs to the sample of \cite{Hsyu+2020} and report $\Yp = 0.2370^{+0.0033}_{-0.0034}$, which is lower than the previous measurements at the $\sim 1\sigma$ level. \cite{Matsumoto+2022} suggest that this low $\Yp$ value can be explained by lepton asymmetry \citep{Kohri+1997, Kawasaki_Murai_2022}, which allows $\neff \sim 3.45$ at the 68\% confidence level and may help alleviate the Hubble tension. 
Interestingly, the slightly lower $\Yp$ value is also reported by the recent CMB measurement with the Atacama Cosmology Telescope (ACT), which suggests $\Yp = 0.227\pm0.014$ at 68\% confidence level \citep{Calabrese+2025}, although the precision of the $\Yp$ value is still insufficient to show the lepton asymmetry beyond the uncertainty.

To determine $\Yp$ precisely, we need to reduce both of the statistical and systematic uncertainties. Previous studies have the large statistical uncertainty due to the small number of EMPGs \citep{Hsyu+2020, Matsumoto+2022}, requiring the substantial extrapolation to the zero metallicity.
To increase the number of EMPGs that can be securely used in the $\Yp$ determination, one should conduct NIR observations targeting the He \textsc{i} $\lambda$ 10830 \AA \, emission line. 
\cite{Peimbert_2007} also argue that the systematic uncertainty arising from the collision strength data of the H \textsc{i} lines represents the largest contribution to the total systematics, with an uncertainty of $\Delta \Yp = \pm0.0015$. Additionally, the slope of the He/H–O/H correlation, which is influenced by the chemical evolution of galaxies, also contributes significantly ($\Delta \Yp = \pm0.0010$; see also \citealt{Fukushima+2024} for the theoretical work on the chemical evolution, which shows uncertainty in the slope by different chemical yield models). It is thus important to increase the number of EMPGs not only to improve statistical precision, but also to reduce systematic uncertainties associated with the chemical evolution.

This work aims to measure $\Yp$ with high precision and to test the validity of previous claims of a low $\Yp$. We present Subaru observations for 29 galaxies, including 14 EMPGs observed in EMPRESS (Extremely Metal-Poor Representatives Explored by the Subaru Survey) 3D project (PI: M. Ouchi), which is an extended program of EMPRESS motivated by its successful results \citep{Kojima+2020, Kojima+2021, Isobe+2021, Isobe+2022, Nakajima+2022, Xu+2022, Umeda+2022, Isobe+2023, Nakajima+2024, Nishigaki+2023, Xu+2024, Watanabe+2024, Hatano+2024}.
This paper is organized as follows. In Section \ref{sec:sample_data}, we describe the sample of galaxies. We present our observations and data reduction in Section \ref{sec:observation}. In Section \ref{sec:analysis}, we explain the flux measurements and abundance determinations. In Section \ref{sec:result}, we derive
the primordial helium abundance, and compare it with the previous measurements. The cosmological implications are discussed in Section \ref{sec:discussion}. In Section \ref{sec:summary}, we summarize this study.

\section{Sample and Data}\label{sec:sample_data}

We use 29 galaxies observed in this work, as well as 58 galaxies whose chemical abundances are determined in literature. These 29 and 58 galaxies are referred to as ``our sample" and ``literature sample", respectively, which are described in the following sections.

\subsection{Our Sample}\label{sec:empress_gal}
Our sample galaxies are selected from the Sloan Digital Sky Survey (SDSS; DR16\footnote[1]{\url{https://skyserver.sdss.org/dr16/en/home.aspx}}) and 
\cite{Thuan_Izotov_2005, Papaderos+2008, Izotov+2012, Izotov+2019, Kojima+2020, Nakajima+2022, Xu+2022, Isobe+2022, Nishigaki+2023}, whose metallicities are known to be $<0.4\,Z_\odot$.  
We observed 29 galaxies that are visible in our observing runs of 2022 November 25 with Subaru/Simultaneous-color Wide-field Infrared Multi-object Spectrograph (SWIMS; \citealt{2014SPIE.9147E..6KM, 2016SPIE.9908E..3UM, 10.1117/12.2310060, 10.1117/12.2560422}), and 2024 Octover 20-21, November 14, and 2025 January 12-13 with Subaru/Multi-Object Infrared Camera and Spectrograph (MOIRCS; \citealt{2006SPIE.6269E..16I, 2008PASJ...60.1347S}). In this paper, we present observations of these new 29 galaxies, which are summarized in Table \ref{tab:targets}.

\subsection{Literature Sample}\label{sec:literature_gal}
We use 54 galaxies from Sample 1 of \cite{Hsyu+2020}, who use a combination of the galaxies from the SDSS and the previous studies \citep{Izotov+2004,Izotov+2007,Izotov+2014}. This is a sample whose observed H \textsc{i} and He \textsc{i} emission line ratios have been successfully reproduced by their photoionization modeling. Most galaxies in this sample are relatively metal-rich, with metallicities exceeding $10\%\,Z_\odot$, while only three EMPGs are included in this sample. We also add 5 EMPGs whose He abundances are reliably determined in \cite{Matsumoto+2022}. However, one out of 5 galaxies, I Zw 18 NW, was also observed in this work. This galaxy is thus excluded from the literature sample. The literature sample consists of $54+4=58$ galaxies, including 7 EMPGs. 

\begin{deluxetable*}{ccccc}
\tablecaption{Subaru galaxies observed in this work \label{tab:targets}}
\tablewidth{0pt}
\tablehead{
\colhead{ID} & \colhead{R.A.} & \colhead{Decl.} & \colhead{$z$} &
\colhead{Optical spectra}
}
\decimalcolnumbers
\startdata
J2115-1734 & 21:15:58.33 & -17:34:45.09 & 0.023 & Magellan/MagE \citep{Kojima+2020} %SDSS_J2115_MagE ID36
\\
J0159+0751 & 01:59:52.75 & +07:51:48.80 & 0.061 & LBT/MODS \citep{Izotov+2017} \\ %ID56
J2302+0049 & 23:02:10.00 & +00:49:38.78 & 0.033 & SDSS \\ %ID28
J2229+2725 & 22:29:33.19 & +27:25:25.60 & 0.076 & LBT/MODS \citep{Izotov+2021} \\ %ID48
J0036+0052 & 00:36:30.40 & +00:52:34.71 & 0.028 & SDSS \\
J0808+1728 & 08:08:40.80 & +17:28:56.49 & 0.044 & SDSS \\
J2136+0414 & 21:36:58.80 & +04:14:04.31 & 0.017 & Magellan/MagE \citep{Nishigaki+2023} \\
J0210-0124 & 02:10:12.07 & -01:24:51.16 & 0.012 & Keck/LRIS \citep{Isobe+2022} \\
J2104-0035 & 21:04:55.30 & -00:35:22.00 & 0.004 & Magellan/MagE (This work) \\
J0159-0622 & 01:59:43.86 & -06:22:32.84 & 0.009 & Keck/LRIS \citep{Isobe+2022} \\
J0107+0103 & 01:07:46.56 & +01:03:52.06 & 0.002 & SDSS \\
J0134-0038 & 01:34:52.00 & -00:38:54.38 & 0.017 & SDSS \\
J0811+4730 & 08:11:52.08 & +47:30:26.24 & 0.044 & SDSS \\
J0845+0131 & 08:45:30.80 & +01:31:51.20 & 0.013 & Magellan/MagE \citep{Xu+2022} \\
J2314+0154 & 23:14:37.55 & +01:54:14.27 & 0.033 & Magellan/MagE \citep{Kojima+2020} \\
J0007+0226 & 00:07:24.49 & +02:26:27.20 & 0.064 & SDSS \\
J0014-0043 & 00:14:34.98 & -00:43:52.03 & 0.013 & SDSS \\
J0226-5017 & 02:26:57.62 & -05:17:47.36& 0.044 & Keck/LRIS \citep{Isobe+2022} \\
J0228-0210 & 02:28:02.59 & -02:10:55.55 & 0.042 & Magellan/MagE (This work) \\
J0248-0817 & 02:48:15.95 & -08:17:16.51 & 0.005 & SDSS \\
SBS0335E & 03:37:44.06 & -05:02:40.19 & 0.014 & VLT/FORS1+UVES \citep{Izotov+2009} \\
J0833+2508 & 08:33:35.65 & +25:08:47.14 & 0.007 & SDSS \\
Mrk996 & 01:25:04.50 & +06:35:07.80 & 0.005 & HST/FOS \citep{Thuan+1996} \\
J0335-0038 & 03:35:26.64 & -00:38:11.33 & 0.023 & SDSS \\
J0301+0114 & 03:01:35.57 & +01:14:20.06 & 0.043 & SDSS \\
J0313+0006 & 03:13:00.05 & +00:06:12.10 & 0.029 & SDSS \\
J0825+1846 & 08:25:40.44 & +18:46:17.20 & 0.038 & SDSS \\
J0815+2156 & 08:15:52.00 & +21:56:23.65 & 0.141 & SDSS \\
I Zw 18 NW & 09:34:02.03 & +55:14:28.07 & 0.002 & MMT/Blue channel \citep{Thuan_Izotov_2005}
\enddata
\tablecomments{(1): ID. (2): Right ascension. (3): Declination. (4): Redshift. (5): Reference for optical spectra.
}
\end{deluxetable*}

\begin{deluxetable*}{cccccc}
\tablecaption{Near-Infrared Spectroscopy \label{tab:observation}}
\tablewidth{0pt}
\tablehead{
\colhead{ID} & \colhead{Instrument} & \colhead{Exp. Time (s)} & \colhead{Observation Date} & \colhead{Seeing ($^{\prime\prime}$)} & \colhead{$F(\mathrm{He \, \textsc{i} \, \lambda10830\AA})/F(\mathrm{P\gamma})$}
}
\decimalcolnumbers
\startdata
J2115-1734 & SWIMS & 900 & 2022 November 25 & 0.6 & $9.54\pm1.49$ \\
J0159+0751 & SWIMS & 900 & 2022 November 25 & 0.6 & $5.43\pm3.62$ \\
J2302+0049 & SWIMS & 1200 & 2022 November 25 & 0.6 &  $1.27\pm0.15$\\
J2229+2725 & SWIMS & 2400 & 2022 November 25 & 0.6 & $4.73\pm0.39$ \\
J0036+0052 & MOIRCS & 2100 & 2024 October 20 & 0.8 & $3.77\pm0.16$ \\
J0808+1728 & MOIRCS & 600 & 2024 October 20 & 0.8 & $4.14\pm0.64$ \\
J2136+0414 & MOIRCS & 1200 & 2024 October 20 & 0.6 & $3.87\pm0.19$ \\
J0210-0124 & MOIRCS & 1200 & 2024 October 20 & 0.8 & $2.86\pm0.18$ \\
J2104-0035 & MOIRCS & 2400 & 2024 October 20 & 0.9 & $\cdots^a$ \\
J0159-0622 & MOIRCS & 1800 & 2024 October 20 & 0.8 & $2.83\pm0.11$ \\
J0107+0103 & MOIRCS & 1200 & 2024 October 20 & 0.8 & $\cdots^a$ \\
J0134-0038 & MOIRCS & 600 & 2024 October 20 & 0.8 & $1.77\pm0.20$ \\
J0811+4730 & MOIRCS & 5400 & 2024 October 20 & 0.8 &  $3.43\pm0.30$\\
J0845+0131 & MOIRCS & 600 & 2024 October 20 & 0.8 & $2.23\pm0.25$ \\
J2314+0154 & MOIRCS & 5400 & 2024 October 21 & 0.7 & $2.58\pm0.68$ \\
J0007+0226 & MOIRCS & 1200 & 2024 October 21 & 0.5 & $3.63\pm0.12$ \\
J0014-0043 & MOIRCS & 600 & 2024 October 21 & 0.4 & $2.52\pm0.10$ \\
J0226-5017 & MOIRCS & 1200 & 2024 October 21 & 0.7 & $3.52\pm0.16$ \\
J0228-0210 & MOIRCS & 1200 & 2024 October 21 & 0.7 & $2.06\pm0.11$ \\
J0248-0817 & MOIRCS & 600 & 2024 October 21 & 0.5 & $2.90\pm0.02$ \\
SBS 0335E & MOIRCS & 600 & 2024 October 21 & 0.5 &  $3.68\pm0.02$\\
J0833+2508 & MOIRCS & 2700 & 2024 October 21 & 0.5 & $2.51\pm0.24$ \\
Mrk 996 & MOIRCS & 450 & 2024 November 14 & 0.9 & $7.99\pm0.18$ \\
J0335-0038 & MOIRCS & 1200 & 2025 January 12 & 0.9 & $3.12\pm0.24$ \\
J0301+0114 & MOIRCS & 1800 & 2025 January 12 & 0.9 & $2.97\pm0.33$ \\
J0313+0006 & MOIRCS & 1800 & 2025 January 12 & 0.9 & $2.50\pm0.28$ \\
J0825+1846 & MOIRCS & 600 & 2025 January 12 & 0.9 & $2.95\pm0.17$ \\
J0815+2156 & MOIRCS & 900 & 2025 January 12 & 0.8 & $3.37\pm0.23$ \\
I Zw 18 NW & MOIRCS & 900 & 2025 January 12 & 1.0 & $1.85\pm0.05$ 
\enddata
\tablecomments{(1) ID. (2) Instruments for our NIR spectroscopy. (3) Total exposure time. (4) Date of our NIR spectroscopy. (5) Seeing size. (6) Observed flux ratio of He \textsc{i} $\lambda$10830\AA\, to P$\gamma$. \\
$^a$Emission lines are not detected.}
\end{deluxetable*}

\section{Observations and Data Reduction}\label{sec:observation}
\subsection{NIR Observations}

Our observations aim to detect He \textsc{i} $\lambda$10830\AA\, and P$\gamma$ emission lines for reliable measurements of He/H. To this end, the 29 galaxies were observed with Subaru/SWIMS or MOIRCS. These observations are summarized in Table \ref{tab:observation}.

\subsubsection{SWIMS Observations}\label{sec:swims}
We conducted NIR spectroscopy with SWIMS long-slit spectroscopy mode for 4 of our sample galaxies, J2302+0049, J2115-1734, J2229+2725, and J0159+0751 on 2022 November 25 (PI: M. Ouchi). SWIMS has blue and red channels with the dichroic at 1.4 $\mu$m, covering the wavelength range of 0.9-2.5 $\mu$m. We used $zJ$ and $HK_\mathrm{s}$ grisms for blue and red channels, 
accomplishing
%resulting in 
the spectral resolutions of $R \sim 700-1200$ and $R \sim 600-1000$, respectively. 
We used an ABBA (ABA) dither pattern for J2302+0049 and J2229+2725 (J2115-1734 and J0159+0751)
with a 300 second exposure at each dither position.
A nearby A0V Hipparcos star, HIP26934, was observed at the end of the observing run for flux calibration.

These data were reduced using the IRAF package. We performed flat-fielding, wavelength calibration, sky subtraction, background subtraction, and cosmic-ray cleaning. Wavelength solutions for the SWIMS spectra were determined with 
%obtained from 
OH sky lines. 
One-dimensional spectra were then extracted from the two-dimensional spectra with a boxcar aperture that encompasses the whole emission line component, with a typical width of 20 pixels. We then performed flux calibration using the observed A0V star. 
Error spectra were extracted with the read-out/photon noise of sky+object emission. The example SWIMS spectrum is shown in Figure \ref{fig:spec}. 

\subsubsection{MOIRCS Observations}
\begin{figure*}
    \centering
    \includegraphics[width=1\linewidth]{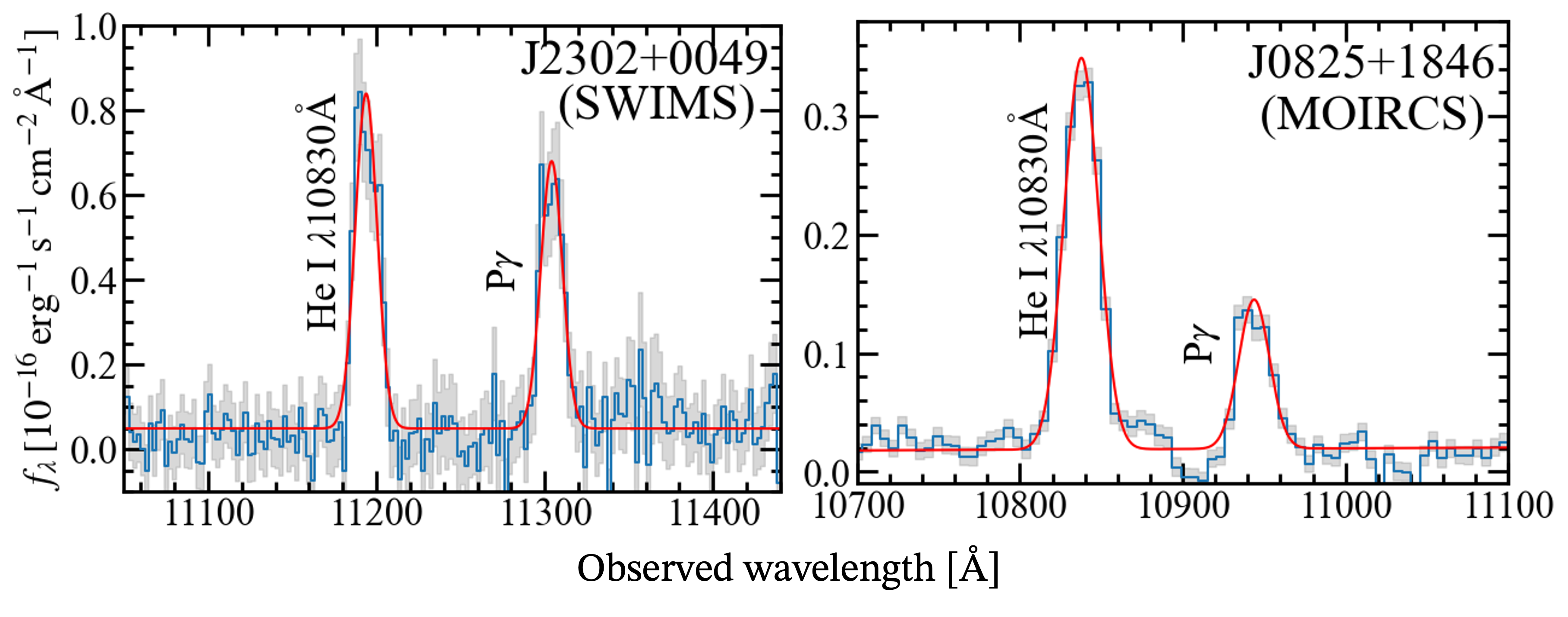}
    \caption{Example NIR spectra of our sample galaxies, J2302+00049 taken with SWIMS and J0825+1846 taken with MOIRCS. The blue histograms and gray shaded regions represent the spectra and their error. The He \textsc{i} 10830 \AA\, and P$\gamma$ emission lines are highlighted. The red lines indicate the best-fit double gaussian.}
    \label{fig:spec}
\end{figure*}

The MOIRCS observations were performed for 25 of our sample galaxies on 2024 October 20-21, November 14, and 2025 January 12-13 (PI: A. Matsumoto), although we did not obtain any science spectrum on 2025 January 13 due to the cloudy weather. 
We used $zJ500$ grism, which covers $0.9-1.8\,\mu\mathrm{m}$ with $R\sim500$. We adopted ABAB dither pattern with an individual exposure time of 150 s. Total exposure times for each galaxy are summarized in Table \ref{tab:observation}. We observed nearby A0V Hipparcos stars (HIP43018, HIP13917, HIP109452, HIP10185, and HIP27848) in each night for flux calibration. The data reduction was performed in the same manner as described in Section \ref{sec:swims} except for the wavelength calibration, which we utilized the ThAr lamp for the MOIRCS observations. The example MOIRCS spectrum is shown in Figure \ref{fig:spec}.

There were two objects that we failed to detect emission lines. The spectrum of J2104-0035 overlapped with a nearby bright star when dithered frames were stacked, making it difficult to extract the spectrum of the object. We did not detect emission lines of J0107+0103 probably due to its extended and diffuse structure. We thus exclude these galaxies from the following analyses, obtaining $29-2=27$ galaxies used in the analysis.

\subsection{Optical Observations}
We use optical spectra taken from previous observations. We use SDSS spectra if available. For galaxies whose SDSS spectra are not available, we use spectra taken in previous EMPRESS observations with Keck/Low Resolution Imaging Spectrograph (LRIS; \citealt{Oke+1995}) or Magellan/Magellan Echellette Spectrograph (MagE; \citealt{Marshall+2008}). We note that for J0159+0751, J2229+2725, SBS 0335E, Mrk 996, and I Zw 18 NW, we use flux measurements taken from %LBT/MODS observations from 
\cite{Izotov+2017}, %VLT/FORS1 and UVES observations from 
\cite{Izotov+2021},
\cite{Izotov+2009}, %HST/FOS observations from 
\cite{Thuan+1996}, or %MMT blue channel spectrograph observations from 
\cite{Thuan_Izotov_2005}, respectively, because of their higher S/N. In the following sections, we briefly describe the LRIS and MagE observations.

\subsubsection{Keck/LRIS Observations}
J0210-0124, J0159-0622, and J0226-5017 were observed with Keck/LRIS on 2019 August 31 (PI: T. Kojima; \citealt{Isobe+2022}). We used the 600 lines $\mathrm{mm^{-1}}$ grism blazed at 4000 \AA\, on the blue channel and the 600 lines $\mathrm{mm^{-1}}$ grating blazed at 7500 \AA\, on the red channel, which accomplish the wavelength coverages of $3000-5500$ and $6000-9000$ \AA\, with the spectral resolutions of $\sim4$ and $5$ \AA\, in FWHM, respectively. Exposure times were 1200 s for all of the three galaxies. Data reduction was performed with IRAF package. We conducted bias subtraction, flat fielding, cosmic ray cleaning, sky subtraction, wavelength calibration, one-dimensional spectrum extraction, flux calibration, atmospheric absorption correction, and Galactic reddening correction. More detailed descriptions of the observations and data reduction can be found in \cite{Isobe+2022}.

\subsubsection{Magellan/MagE Observations}
J2115-1734, J2314+0154, J0845+0131, J2136+0414, J2104-0035, and J0228-0210 were observed with Magellan/MagE during observing runs of 2018 June 13, 2021 February 9, July 10, and October 9 (PI: M. Rauch). Details of these observations are described in \cite{Kojima+2020}, \cite{Xu+2022}, and \cite{Nishigaki+2023}.\footnote[2]{Although the observations of J2104-0035 and J0228-0817 are described in \cite{Nishigaki+2023}, these galaxies have not been included in their sample.} The wavelength coverage of the MagE spectroscopy was $3100-10000$ \AA\,, with a spectral resolution of $R\sim4000$. Exposure times were $300-3600$ s depending on the luminosity of the target. The raw data of J2115-1734 and J2314+0154 were reduced with the MagE pipeline from the Carnegie Observatories Software Repository.\footnote[3]{\url{https://code.obs.carnegiescience.edu}} The bias subtraction, flat fielding, scattered-light subtraction, 2D spectrum subtraction, sky subtraction, wavelength calibration, cosmic ray removal, and 1D-spectrum extraction were conducted with the MagE pipeline. We then conducted flux calibration with the standard star Feige 110 using IRAF routines. The details of these data reduction are described in \cite{Kojima+2020}. The other MagE data were reduced with PypeIt \citep{Prochaska+2020a, Prochaska+2020b}. We performed flat fielding, wavelength calibration, sky subtraction, cosmic ray removal, 1D-spectrum extraction, and flux calibration. Because PypeIt did not fit a background model correctly around bright extended emission lines such as H$\alpha$ and [O \textsc{iii}] $\lambda$5007, we %adopted flat sky background in pixels where strong emission lines exist and 
estimated the flux of flat sky background by averaging sky background in the nearby pixels. The details of these data reduction are described in \cite{Xu+2022}.

\section{Analysis}\label{sec:analysis}
\subsection{Flux and EW Measurements}\label{sec:flux}
We measure the hydrogen, helium, oxygen, nitrogen, and sulfur emission line fluxes and their equivalent widths (EWs) of our sample galaxies to derive the chemical abundances. We measure the line fluxes and EWs of \textsc{[O ii]} $\lambda\lambda$3727,3729,  He \textsc{i} $\lambda$4026, \textsc{[O iii]} $\lambda$4363, He \textsc{i} $\lambda$4471, He \textsc{ii} $\lambda$4686, \textsc{[O iii]} $\lambda\lambda$4959,5007, He \textsc{i} $\lambda$5015, He \textsc{i} $\lambda$5876, [N \textsc{ii}]$\lambda\lambda$ 6548,6583, He \textsc{i} $\lambda$6678, \textsc{[S ii]} $\lambda\lambda$6717,6731, He \textsc{i} $\lambda$7065, \textsc{[O ii]} $\lambda\lambda$7320,7330, \textsc{[S iii]} $\lambda$9069, He \textsc{i} $\lambda$10830, P$\gamma$, the Balmer series from H$\alpha$ to H$\delta$, and the blended H8+He \textsc{i} $\lambda$3889 by fitting the Gaussian profile and linear continuum. Note that we fit multiple Gaussian profiles to the \textsc{[O iii]} $\lambda$5007+He \textsc{i} $\lambda$5015, H$\alpha$+[N \textsc{ii}]$\lambda\lambda$ 6548,6583, He \textsc{i} $\lambda$10830+P$\gamma$ lines and doublets. Some emission lines have broad absorption profiles. For these lines, we exclude the absorption profiles from the fitting ranges. We conduct Markov Chain Monte Carlo (MCMC) technique to obtain the best-fit flux and its error by taking the median and 68\% confidence level with {\tt\string emcee} package \citep{Foreman-Mackey+2013}.

To minimize possible slit-loss effect arising from uncertainties in fiber positions of the optical and NIR spectroscopy, we normalize the NIR lines ([S \textsc{iii}] $\lambda$9069 and He \textsc{i} $\lambda$10830) with the P$\gamma$ line and renormalize them with the H$\beta$ line using the Case B theoretical $\mathrm{P\gamma/H\beta}$ flux ratio.

The NIR He {\sc i} 10830 \AA/P$\gamma$ line ratio is summarized in Table \ref{tab:observation}. Some of the galaxies (J2302+0049, J0134-0038, and I Zw 18 NW) show very low line ratios of $<2$, indicative of low electron densities. Although such low line ratios are rare, comparable values have been reported in previous studies (e.g., \citealp{Hsyu+2020, Matsumoto+2022}).

\begin{deluxetable}{cc}
\tablecaption{References for data used in {\tt\string YMCMC} \label{tab:atomic data}}
\tablewidth{0pt}
\tablehead{
\colhead{Data} & \colhead{Reference}
}
\startdata
H \textsc{i} emissivity & \cite{Storey_Sochi_2015} \\
He \textsc{i} emissivity & \cite{Aver+2013} \\
H \textsc{i} collisional excitation correction & \cite{Anderson+2000, Anderson+2002}, \cite{Omidvar_1983}, \cite{Hsyu+2020} \\
Absorption correction & \cite{Aver_2015} \\
Reddening correction & \cite{Cardelli+1989} \\
Optical depth function & \cite{Benjamin+2002}, \cite{Aver_2015}
\enddata
\end{deluxetable}

\begin{deluxetable*}{cccccccccc}
\tablewidth{12pt}
\tablecaption{Best recovered parameters from MCMC analysis}
\label{tab:mcmc_recovered_params}
\tablehead{ 
\colhead{ID} & \colhead{$y^{+}$} & \colhead{$T_{\rm e}$} & \colhead{log$_{10}(n_{\rm e}/\rm cm^{-3}$)} & \colhead{$c$(H$\beta$)} & \colhead{$a_{\rm H}$} & \colhead{$a_{\rm He}$} & \colhead{$\tau_{\rm He}$} & \colhead{log$_{10}(\xi$)} & \colhead{$\chi^2$} \\
\colhead{} & \colhead{} & \colhead{[K]} & \colhead{} & \colhead{} & \colhead{[\AA]} & \colhead{[\AA]} & \colhead{} & \colhead{} & \colhead{}
}
\startdata
J2115-1734 & $\err{0.0740}{0.0031}{0.0022}$ & $\err{19052}{1508}{1284}$ & $\err{2.79}{0.08}{0.10}$ & $\err{0.28}{0.01}{0.01}$  & $\err{0.09}{0.14}{0.07}$ & $\err{0.15}{0.16}{0.10}$ & $\err{0.11}{0.24}{0.09}$ & $\err{-4.87}{0.83}{0.76}$ & 755 \\
J0159+0751 & $\err{0.0719}{0.0022}{0.0019}$ & $\err{15304}{1433}{1027}$ & $\err{2.98}{0.02}{0.04}$ & $\err{0.005}{0.009}{0.003}$ & $\err{0.39}{0.58}{0.29}$ & $\err{0.02}{0.03}{0.01}$ & $\err{4.76}{0.18}{0.34}$ & $\err{-4.44}{1.14}{1.07}$ & 96 \\
J2302+0049 & $\err{0.0723}{0.0030}{0.0029}$ & $\err{12311}{1870}{1489}$ & $\err{0.39}{0.39}{0.27}$ & $\err{0.14}{0.03}{0.03}$ & $\err{0.39}{0.48}{0.28}$ & $\err{0.15}{0.19}{0.11}$ & $\err{1.39}{1.08}{0.87}$ & $\err{-3.57}{1.84}{1.68}$ & 23 \\
%J2229+2725 & $\err{0.0935}{0.0217}{0.0122}$ & $\err{20469}{1105}{1881}$ & $\err{2.34}{0.14}{0.16}$ & $\err{0.26}{0.07}{0.14}$ & $\err{3.08}{3.42}{2.15}$ & $\err{1.85}{1.15}{1.09}$ & $\err{3.54}{1.02}{1.65}$ & $\err{-3.20}{0.67}{1.89}$ & 1.79\\
J2229+2725 & $\err{0.0731}{0.0037}{0.0034}$ & $\err{21365}{473}{974}$ & $\err{2.26}{0.12}{0.13}$ & $\err{0.01}{0.01}{0.00}$ & $\err{1.47}{2.19}{1.06}$ & $\err{0.85}{0.93}{0.59}$ & $\err{4.27}{0.48}{0.62}$ & $\err{-4.97}{0.86}{0.72}$ & 28 \\
J0036+0052 & $\err{0.0896}{0.0079}{0.0065}$ & $\err{17528}{2166}{2103}$ & $\err{2.02}{0.16}{0.18}$ & $\err{0.05}{0.04}{0.03}$ & $\err{1.97}{0.86}{0.86}$ & $\err{0.32}{0.38}{0.24}$ & $\err{0.59}{0.80}{0.42}$ & $\err{-4.18}{1.09}{1.23}$ & 3.6 \\
J0808+1728 & $\err{0.0926}{0.0170}{0.0122}$ & $\err{16439}{2088}{2162}$ & $\err{2.02}{0.35}{0.84}$ & $\err{0.14}{0.06}{0.10}$ & $\err{0.85}{0.99}{0.61}$ & $\err{0.61}{0.72}{0.43}$ & $\err{1.17}{1.60}{0.87}$ & $\err{-2.94}{0.88}{2.05}$ & 4.3\\
J2136+0414 & $\err{0.0925}{0.0051}{0.0040}$ & $\err{12604}{1804}{1480}$ & $\err{2.28}{0.16}{0.15}$ & $\err{0.28}{0.02}{0.05}$ & $\err{4.28}{2.29}{1.97}$ & $\err{3.39}{0.47}{0.94}$ & $\err{0.31}{0.43}{0.23}$ & $\err{-3.12}{1.65}{1.92}$ & 16 \\
J0210-0124 & $\err{0.1321}{0.0124}{0.0221}$ & $\err{15459}{1883}{1777}$ & $\err{0.57}{0.48}{0.39}$ & $\err{0.08}{0.07}{0.05}$ & $\err{0.40}{0.47}{0.29}$ & $\err{0.07}{0.08}{0.05}$ & $\err{1.18}{0.77}{0.68}$ & $\err{-1.63}{0.68}{1.69}$ & 910 \\
%J2104-0035 & &&&&&&&\\
J0159-0622 & $\err{0.0902}{0.0019}{0.0020}$ & $\err{13595}{1027}{927}$ & $\err{1.62}{0.20}{0.29}$ & $\err{0.02}{0.01}{0.01}$ & $\err{0.76}{0.54}{0.46}$ & $\err{0.77}{0.17}{0.17}$ & $\err{3.19}{0.36}{0.36}$ & $\err{-4.39}{1.23}{1.09}$ & 28 \\
J0134-0038 & $\err{0.0829}{0.0057}{0.0057}$ & $\err{11008}{1106}{692}$ & $\err{0.59}{0.50}{0.41}$ & $\err{0.15}{0.04}{0.04}$ & $\err{0.72}{0.86}{0.52}$ & $\err{0.23}{0.33}{0.17}$ & $\err{1.21}{1.47}{0.87}$ & $\err{-3.22}{1.90}{1.90}$ & 6.9  \\
J0811+4730 & $\err{0.0932}{0.0078}{0.0078}$ & $\err{15044}{1540}{1590}$ & $\err{1.88}{0.26}{0.49}$ & $\err{0.01}{0.01}{0.01}$ & $\err{0.40}{0.63}{0.29}$ & $\err{0.34}{0.51}{0.26}$ & $\err{2.46}{1.28}{1.29}$ & $\err{-4.30}{1.15}{1.10}$ & 17 \\
J0845+0131 & $\err{0.1026}{0.0030}{0.0032}$ & $\err{10549}{866}{407}$ & $\err{0.52}{0.48}{0.35}$ & $\err{0.37}{0.02}{0.02}$ & $\err{1.66}{1.17}{0.99}$ & $\err{0.22}{0.35}{0.17}$ & $\err{0.86}{0.69}{0.55}$ & $\err{-3.15}{1.95}{1.98}$ & 115 \\
J2314+0154 & $\err{0.0773}{0.0057}{0.0057}$ & $\err{15309}{2201}{2278}$ & $\err{1.23}{0.70}{0.83}$ & $\err{0.26}{0.04}{0.05}$ & $\err{6.66}{1.96}{2.52}$ & $\err{3.05}{0.69}{1.21}$ & $\err{3.81}{0.88}{1.75}$ & $\err{-4.13}{1.34}{1.25}$ & 3.8 \\
J0007+0226 & $\err{0.0910}{0.0083}{0.0063}$ & $\err{16414}{1784}{1772}$ & $\err{2.05}{0.14}{0.15}$ & $\err{0.06}{0.04}{0.05}$ & $\err{0.87}{1.18}{0.63}$ & $\err{1.21}{1.03}{0.78}$ & $\err{1.99}{0.76}{0.76}$ & $\err{-2.81}{0.71}{1.94}$ & 5.8 \\
J0014-0043 & $\err{0.0864}{0.0071}{0.0043}$ & $\err{11987}{1395}{1178}$ & $\err{1.58}{0.31}{0.66}$ & $\err{0.16}{0.03}{0.08}$ & $\err{0.63}{0.63}{0.42}$ & $\err{0.29}{0.23}{0.18}$ & $\err{0.89}{0.67}{0.56}$ & $\err{-2.22}{1.34}{2.50}$ & 5.5 \\
J0226-5017 & $\err{0.1230}{0.0078}{0.0098}$ & $\err{17231}{2293}{2246}$ & $\err{0.70}{0.60}{0.48}$ & $\err{0.02}{0.03}{0.02}$ & $\err{2.60}{0.30}{0.32}$ & $\err{0.25}{0.29}{0.18}$ & $\err{2.39}{1.72}{1.71}$ & $\err{-4.68}{1.01}{0.89}$ & 9.2\\
J0228-0210 & $\err{0.1253}{0.0193}{0.0610}$ & $\err{15092}{2975}{3676}$ & $\err{1.28}{0.56}{0.77}$ & $\err{0.26}{0.24}{0.12}$ & $\err{0.62}{1.31}{0.49}$ & $\err{0.58}{0.69}{0.42}$ & $\err{1.73}{1.77}{1.20}$ & $\err{-1.40}{0.74}{3.53}$ & 276 \\
J0248-0817 & $\err{0.0871}{0.0015}{0.0012}$ & $\err{14097}{635}{659}$ & $\err{1.86}{0.06}{0.07}$ & $\err{0.24}{0.01}{0.02}$ & $\err{0.27}{0.43}{0.20}$ & $\err{0.12}{0.12}{0.08}$ & $\err{0.08}{0.12}{0.06}$ & $\err{-4.03}{1.40}{1.35}$ & 42 \\
SBS0335E & $\err{0.0731}{0.0024}{0.0026}$ & $\err{15558}{876}{730}$ & $\err{2.32}{0.06}{0.05}$ & $\err{0.01}{0.01}{0.01}$ & $\err{0.03}{0.03}{0.02}$ & $\err{0.02}{0.01}{0.01}$ & $\err{4.96}{0.03}{0.07}$ & $\err{-4.50}{0.93}{0.95}$ & 273 \\
J0833+2508 & $\err{0.0929}{0.0123}{0.0113}$ & $\err{14994}{4482}{3407}$ & $\err{0.76}{0.70}{0.53}$ & $\err{0.15}{0.08}{0.09}$ & $\err{0.34}{0.50}{0.26}$ & $\err{0.16}{0.26}{0.12}$ & $\err{2.53}{1.61}{1.65}$ & $\err{-3.13}{1.33}{1.78}$ & 440 \\ %7.82 \\
Mrk 996 & $\err{0.1336}{0.0111}{0.0141}$ & $\err{13590}{2615}{2046}$ & $\err{2.62}{0.13}{0.13}$ & $\err{0.48}{0.01}{0.02}$ & $\err{0.09}{0.12}{0.07}$ & $\err{0.23}{0.12}{0.11}$ & $\err{4.45}{0.43}{0.64}$ & $\err{-2.01}{0.96}{1.34}$ & 82 \\
J0335-0038 & $\err{0.0899}{0.0080}{0.0055}$ & $\err{14489}{2031}{1879}$ & $\err{1.76}{0.31}{0.62}$ & $\err{0.27}{0.03}{0.04}$ & $\err{1.47}{0.83}{0.73}$ & $\err{0.51}{0.52}{0.34}$ & $\err{2.67}{1.54}{1.64}$ & $\err{-3.69}{1.54}{1.57}$ & 6.4 \\
J0301+0114 & $\err{0.0980}{0.0087}{0.0066}$ & $\err{14000}{1649}{1665}$ & $\err{1.17}{0.62}{0.81}$ & $\err{0.22}{0.03}{0.05}$ & $\err{1.67}{1.23}{0.97}$ & $\err{0.75}{0.92}{0.53}$ & $\err{1.65}{1.55}{1.14}$ & $\err{-3.46}{1.55}{1.75}$ & 2.5 \\
%J0313+0006 & $\err{0.1188}{0.0222}{0.0502}$ & $\err{16189}{2328}{1840}$ & $\err{2.04}{0.15}{0.17}$ & $\err{0.18}{0.12}{0.11}$ & $\err{2.50}{1.56}{1.47}$ & $\err{0.12}{0.18}{0.09}$ & $\err{0.72}{0.88}{0.52}$ & $\err{-1.30}{0.70}{1.88}$ & 34 \\
J0313+0006 & $\err{0.0645}{0.0052}{0.0032}$ & $\err{17173}{2094}{2008}$ & $\err{1.74}{0.35}{0.74}$ & $\err{0.31}{0.10}{0.11}$ & $\err{3.16}{1.71}{1.32}$ & $\err{0.10}{0.15}{0.08}$ & $\err{0.46}{0.65}{0.34}$ & $\err{-4.30}{1.40}{1.18}$ & 21
\\
J0825+1846 & $\err{0.0737}{0.0031}{0.0029}$ & $\err{10607}{832}{441}$ & $\err{2.39}{0.15}{0.15}$ & $\err{0.11}{0.02}{0.02}$ & $\err{9.59}{0.30}{0.52}$ & $\err{2.10}{0.37}{0.35}$ & $\err{4.85}{0.11}{0.21}$ & $\err{-2.11}{1.48}{2.59}$ & 86 \\
J0815+2156 & $\err{0.0871}{0.0055}{0.0047}$ & $\err{14821}{1923}{1836}$ & $\err{2.02}{0.20}{0.24}$ & $\err{0.09}{0.03}{0.03}$ & $\err{1.81}{1.26}{1.10}$ & $\err{0.90}{0.57}{0.49}$ & $\err{1.69}{1.34}{1.06}$ & $\err{-3.87}{1.36}{1.43}$ & 3.4 \\
I Zw 18 NW & $\err{0.0684}{0.0033}{0.0024}$ & $\err{20004}{1294}{1682}$ & $\err{0.35}{0.36}{0.25}$ & $\err{0.05}{0.02}{0.03}$ & $\err{0.10}{0.14}{0.07}$ & $\err{0.15}{0.04}{0.04}$ & $\err{0.20}{0.29}{0.15}$ & $\err{-3.62}{0.61}{1.57}$ & 8.2 \\
\enddata
\end{deluxetable*}

\begin{figure*}[tb]
    \centering
    \includegraphics[keepaspectratio, width=\linewidth]{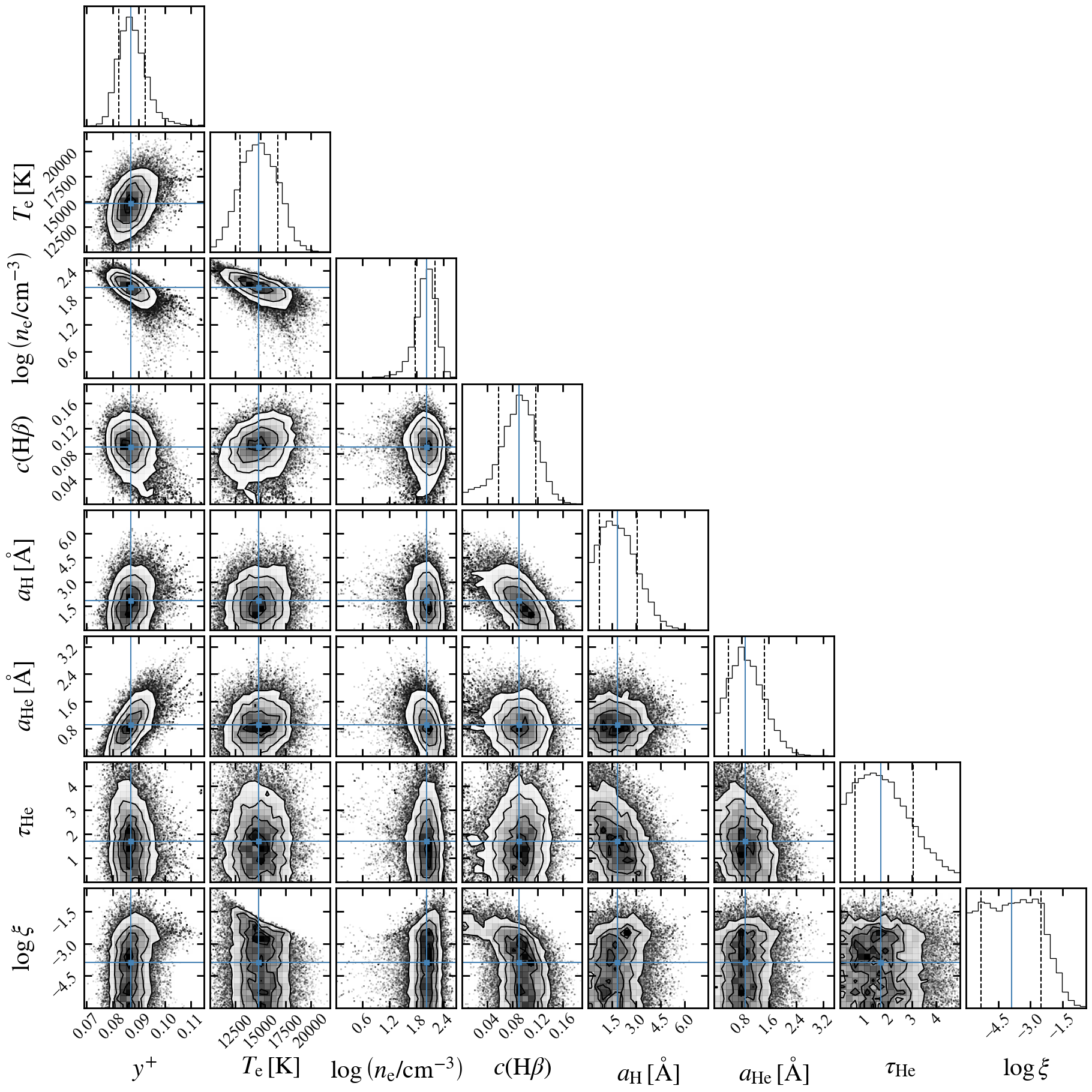}
    \caption{Probability distribution functions (PDFs) of the model parameters for J0815+2156 obtained with {\tt\string YMCMC}. The one and two dimensional PDFs are shown in the diagonal and off-diagonal panels, respectively. 
    The contours denote the 1$\sigma$, 2$\sigma$, and 3$\sigma$ confidence levels. The blue solid and black dashed lines indicate the best-fit values and the 68\% confidence levels, respectively. 
    }
    \label{fig:ymcmc_example}
\end{figure*}

\subsection{Oxygen Abundance}
The oxygen abundance in the H \textsc{ii} region is the sum of the singly and doubly ionized oxygen abundances:
\begin{equation}
    \frac{\mathrm{O}}{\mathrm{H}}\,=\,\frac{\mathrm{O^{+}}}{\mathrm{H^{+}}}\,+\,\frac{\mathrm{O^{++}}}{\mathrm{H^{+}}}, 
\end{equation}
where the neutral oxygen and the triply (or higher-order) ionized oxygen are negligible.
We calculate the $\mathrm{O^{++}}$ abundances using the [O \textsc{iii}] $\lambda\lambda$4959,5007/H$\beta$ flux ratios and the electron temperature of [O \textsc{iii}], $T_\mathrm{e}(\mathrm{O \,\textsc{iii}})$, derived from the [O \textsc{iii}]$\lambda\lambda$4959,5007/[O \textsc{iii}]$\lambda$4363 flux ratio (i.e., the direct temperature method).
The $\mathrm{O^{+}}$ abundances are calculated from the [O \textsc{ii}] $\lambda\lambda$3727,3729/H$\beta$ flux ratios. To obtain the $\mathrm{O^{+}}$ abundance from these [O \textsc{ii}] lines, we need the $\mathrm{O^{+}}$ electron density $n_\mathrm{e}(\mathrm{O\,\textsc{ii}})$ and temperature $T_\mathrm{e}(\mathrm{O\,\textsc{ii}})$. 
We assume that $n_\mathrm{e}(\mathrm{O\,\textsc{ii}})$ is equal to the electron density of the $\mathrm{S^{+}}$ region, $n_\mathrm{e}(\mathrm{S} \, \textsc{ii})$, that is calculated from the [S \textsc{ii}] $\lambda\lambda$6717,6731 doublet.
We estimate $T_\mathrm{e}(\mathrm{O\,\textsc{ii}})$ from $T_\mathrm{e}(\mathrm{O\,\textsc{iii}})$ following the relation from \cite{Pagel+1992}:

\begin{equation}   
T_\mathrm{e}(\mathrm{O\,\textsc{ii}})\,=\,20,000~\mathrm{K} \Big/ \Big(\frac{10,000~\mathrm{K}}{T_\mathrm{e}(\mathrm{O\,\textsc{iii}})}\,+\,0.8\Big).
\end{equation}
The electron temperatures, densities, and oxygen abundances are estimated with {\tt\string PyNeb} \citep{Luridiana+2015}. The total oxygen abundances are shown in Table \ref{table:abundance}. 

\subsection{Helium Abundance}
The number abundance ratio of helium to hydrogen $\mathrm{He/H} \equiv y$ is given by the sum of the abundance ratios of neutral $y^{0}$, singly ionized $y^{+}$, and doubly ionized $y^{++}$ helium to hydrogen:

\begin{equation}
    y = y^{0} + y^{+} + y^{++} 
\end{equation}
We explain the procedures to derive $y^{++}$, $y^{+}$, and $y^{0}$ in the following sections.

\subsubsection{Doubly Ionized Helium}\label{sec:y++}
The $y^{++}$ values are calculated from Equation (17) of \cite{Pagel+1992}:

\begin{equation}
    y^{++} = 0.084 \, \left[\frac{T_\mathrm{e}(\mathrm{O\,\textsc{iii}})}{10^4\,\mathrm{K}}\right]^{0.14}\frac{F\left(\textrm{He\,\textsc{ii}}\,\lambda4686\right)}{F\left(\textrm{H}\beta\right)} ,
\end{equation}
where $F(\lambda)$ is the flux value of the emission line $\lambda$. If there is no detectable He \textsc{ii} $\lambda$4686 line, the $y^{++}$ abundance of the galaxy is assumed to be negligible. The $y^{++}$ values of our sample galaxies are shown in Table \ref{table:abundance}.

\subsubsection{Singly Ionized Helium}\label{sec:y+}
We estimate the $y^{+}$ values of our sample galaxies using the {\tt\string YMCMC} code developed by \cite{Hsyu+2020}. Utilizing the Markov Chain Monte Carlo (MCMC) algorithm, the {\tt\string YMCMC} code solves for the best-fit parameters that reproduce the observed H \textsc{i} and He \textsc{i} emission lines measured in Section \ref{sec:flux}. The free parameters are:
\begin{itemize}
    \item the singly ionized helium abundance, i.e.  $y^{+}$
    \item the electron temperature of the $y^{+}$ region, $T_\mathrm{e}$ [K]
    \item the electron density, $n_\mathrm{e}$ [cm$^{-3}$]
    \item the reddening correction parameter, $c(\mathrm{H\beta})$
    \item the stellar absorption of hydrogen, $a_\mathrm{H}$ [\AA], normalized to the amount of absorption at H$\beta$
    \item the stellar absorption of helium, $a_\mathrm{He}$ [\AA], normalized to the amount of absorption at He \textsc{i}~$\lambda$4471
    \item the helium optical depth parameter, $\tau_\mathrm{He}$, normalized to the value at He \textsc{i}~$\lambda$3889
    % See Benjamin, Skillman, SMits 2002 about the normalization to HeI3889
    \item the ratio of neutral $n({\textrm H\,\textsc{i}})$ to ionized $n({\textrm H\,\textsc{ii}})$ hydrogen number density, $\xi\,\equiv\,n(\textrm{H\,\textsc{i}})/n(\textrm{H\,\textsc{ii}})$. 
\end{itemize}

Using these parameters, the hydrogen flux ratios relative to H$\beta$ are calculated by:

\begin{align}
    \notag
    \frac{F(\lambda)}{F(\mathrm{H}\beta)} = \frac{E(\lambda)}{E(\mathrm{H}\beta)}\frac{\frac{EW(\mathrm{H}\beta)+a_\mathrm{H}(\mathrm{H}\beta)}{EW(\mathrm{H}\beta)}}{\frac{EW(\lambda)+a_\mathrm{H}(\lambda)}{EW(\lambda)}} \\ 
    \times \frac{1+\frac{C}{R}(\lambda)}{1+\frac{C}{R}(\mathrm{H}\beta)}~10^{-f(\lambda)c(\mathrm{H}\beta)},
\end{align}
where $E(\lambda)$, $EW(\lambda)$, $C/R(\lambda)$, and $f(\lambda)$ denotes the emissivity, equivalent width, collisional-to-recombination correction factor, and reddening law for a line $\lambda$. The helium flux ratios relative to H$\beta$ are calculated by:

\begin{align}
    \frac{F(\lambda)}{F(\mathrm{H}\beta)} = y^{+}\frac{E(\lambda)}{E(\mathrm{H}\beta)}\frac{\frac{EW(\mathrm{H}\beta)+a_\textbf{He}(\mathrm{H}\beta)}{EW(\mathrm{H}\beta)}}{\frac{EW(\lambda)+a_\textbf{He}(\lambda)}{EW(\lambda)}}f_\tau(\lambda)
    %\times \frac{1+\frac{C}{R}(\lambda)}{1+\frac{C}{R}(\mathrm{H}\beta)}
    ~10^{-f(\lambda)c(\mathrm{H}\beta)},
\end{align}
where $f_\tau(\lambda)$ is the helium optical depth function. We set $C/R(\lambda)=0$ because the collisional-to-recombination correction for helium is included in the emissivity \citep{Aver+2013}. Atomic data, reddening law, and optical depth function used in our MCMC analysis are summarized in Table \ref{tab:atomic data}.

Following \cite{Hsyu+2020}, we use the flat priors of
\begin{equation}\label{eq:prior}
\begin{split}
    &0.06 \leq y^+ \leq 0.15 \\
    &0 \leq \log_{10}(n_e) \leq 3 \\
    &0 \leq c(\mathrm{H}\beta) \leq 0.5 \\
    &0 \leq a_\mathrm{H} \leq 10 \\
    &0 \leq a_\mathrm{He} \leq 5 \\
    &0 \leq \tau_\mathrm{He} \leq 5 \\
    &-6 \leq \log_{10}(\xi) \leq -0.1.
\end{split}
\end{equation}
Note that we increase the upper boundary of $y^+$ up to 0.15 from that of \cite{Hsyu+2020}, motivated by potential He overabundance found at high-$z$ galaxies \citep{Yanagisawa+2024b}. 
In Section \ref{sec:Yp}, we assess possible bias introduced by the choice of the priors described in \cite{Matsumoto+2022}.

In our {\tt\string YMCMC} analyses, we use 500 walkers and 1000 steps, 800 out of which include burn-in procedures. The best-fit parameters are shown in Table \ref{tab:mcmc_recovered_params}. The best-fit parameters are taken as the median values, with uncertainties defined by the 16–84th percentile range. Figure \ref{fig:ymcmc_example} presents the contours and the histograms for the recovered model parameters of J0815+2156. We find that the degeneracy between $T_\mathrm{e}$ and $n_\mathrm{e}$ is resolved by the He \textsc{i} $\lambda$10830\AA\, line. We also find that the posterior distribution of $\log \xi$ shows a plateau at $\log \xi \lesssim -3$. This occurs because, at such a small value of $\log \xi$ (i.e., a low neutral hydrogen fraction), the contribution of collisional excitation to hydrogen line intensities is so small that we are unable to distinguish between $\log \xi \sim -3$ and lower values. However, this does not significantly affect the posterior distributions of the other parameters.

There is a correlation between $y^+$ and $a_\mathrm{He}$ because, for a given He \textsc{i}/H \textsc{i} flux ratio, a larger $a_\mathrm{He}$ (i.e., stronger He \textsc{i} line absorption) corresponds to an intrinsically higher He \textsc{i}/H \textsc{i} flux ratio, which in turn leads to a larger $y^+$.

\begin{figure}
    \centering
    \includegraphics[width=1\linewidth]{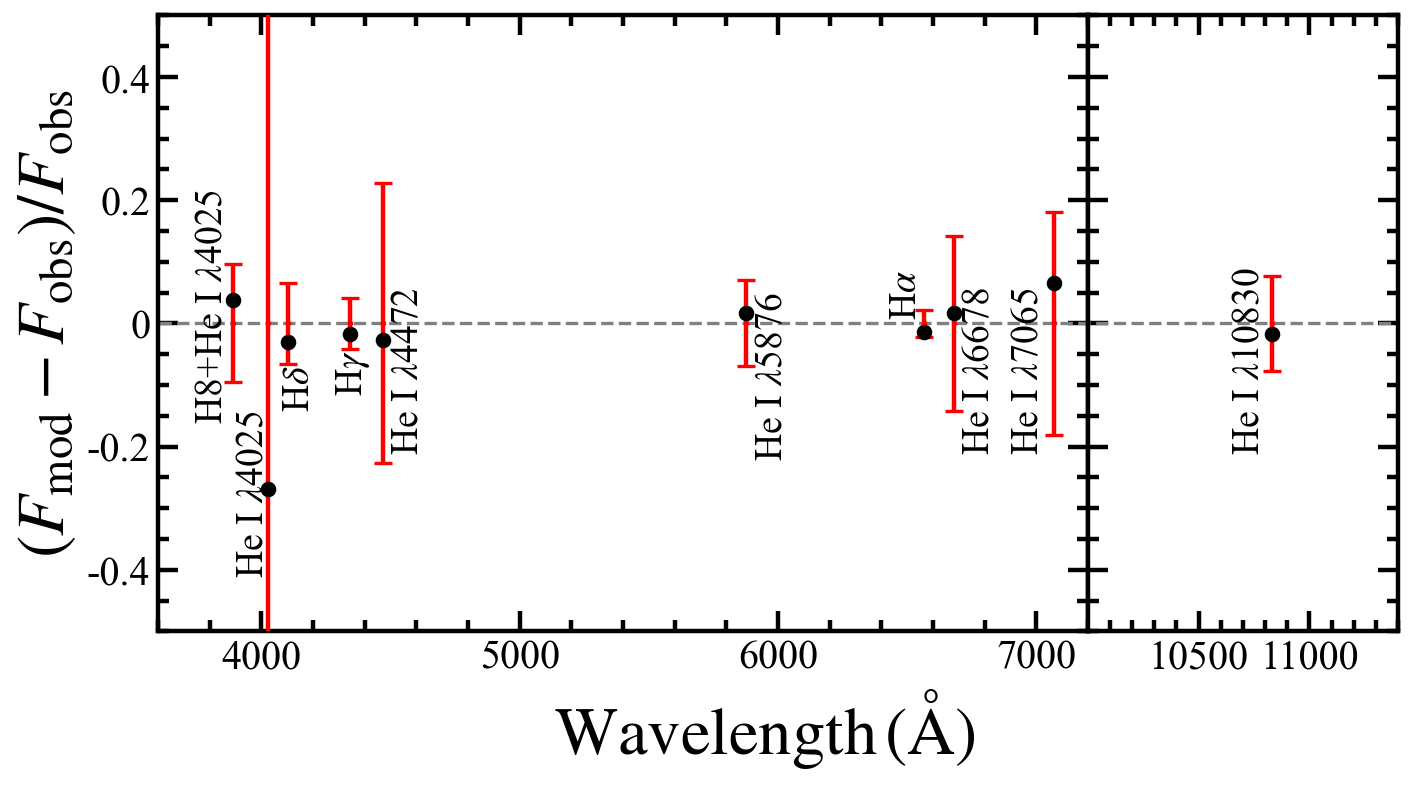}
    \caption{Comparison of the recovered and observed flux ratios of J0014-0043. The black dots represent the relative error of the recovered flux ratios compared to the observations, while the red error bars show the observational $2\sigma$ errors.}
    \label{fig:match}
\end{figure}

We select galaxies whose $y^+$ values are reliably determined with {\tt\string YMCMC}. Following \cite{Hsyu+2020}, we require that all detected He \textsc{i} and H \textsc{i} emission lines are reproduced by the best-fit parameters within the $2\sigma$ levels. We present the example recovered emission line ratios of J0014-0043 in Figure \ref{fig:match}, where we find all detected lines are explained by the model. We find that 11 out of 27 galaxies qualify this criterion. The rest of the 16 galaxies are not well reproduced by {\tt\string YMCMC} model. We thus exclude these 16 galaxies from $\Yp$ determination (note that relaxing this selection criterion only marginally affects the $\Yp$ value). Since {\tt\string YMCMC} assumes simple interstellar medium properties, it may not account for some physical conditions in these galaxies \citep{Hsyu+2020}. For example, it may not reproduce systems that are poorly described by a one-zone temperature model, which {\tt\string YMCMC} assumes.

\subsubsection{Neutral Helium}\label{sec:y0}
Although 
almost all helium atoms are ionized in an H \textsc{ii} region, there may remain some neutral helium atoms because of the difference in the ionization potential energies of He$^+$ (24.6 eV) and H$^+$ (13.6 eV). If the number of He$^+$ ionizing photons is sufficiently large compared to that of H$^+$ ionizing photons (i.e., hard ionizing radiation field), the contribution of neutral helium is expected to be small.
To examine whether the contribution of $y^{0}$ to $y$ is significant or not, we follow \cite{Vilchez+1988} and calculate the radiation softness parameter, $\eta_\mathrm{rds}$, defined as

\begin{equation}
    \eta_\mathrm{rds} = (\mathrm{S}^{++}/\mathrm{S}^+)/(\mathrm{O}^{++}/\mathrm{O}^+).
\end{equation}
We calculate $\mathrm{S}^+$, and $\mathrm{S}^{++}$ with emission lines [S \textsc{ii}] $\lambda\lambda6717,6731$ and [S \textsc{iii}] $\lambda9069$ by using {\tt\string PyNeb}. 
We show the $\eta_\mathrm{rds}$ values in Table \ref{table:abundance}. \cite{Pagel+1992} suggest that the abundance of neutral helium is negligible for a galaxy with $\log \eta_\mathrm{rds} \lesssim 0.9$ based on their photoionization modeling. As J0134-0038, J0226-5017, and J0833-2058 do not meet this criterion, we exclude these three galaxies from our final sample for $\Yp$ determination.

\subsection{Nitrogen Abundance}\label{sec:nitrogen}
\cite{Yanagisawa+2024b} present that the nitrogen-rich ($\log (\mathrm{N/O}) \gtrsim -1.0$) galaxies (hereafter N-rich galaxies) found at $z\sim6$ by James Webb Space Telescope (JWST) observations show extremely strong He \textsc{i} emission, suggesting a possibility of high He/H. The N-rich galaxies may have the chemical compositions that significantly deviate from the primordial abundance,
even though their metallicities are low. To avoid contamination from such galaxies, we exclude galaxies whose $\log (\mathrm{N/O})$ values are higher than $-1.0$. 
We estimate $\mathrm{N^{+}/H^{+}}$ from [N \textsc{ii}] $\lambda\lambda$6548,6583 lines using {\tt\string PyNeb}. Following \cite{Izotov+2006}, we adopt ionization correction factors (ICFs) to derive total N/H from the $\mathrm{N^{+}/H^{+}}$ values. We then obtain N/O by dividing N/H with O/H. The N/O measurements are listed in Table \ref{table:abundance}. Among our sample galaxies, there are 3 galaxies (J0811+4730, J0845+0131, and Mrk 996) applying to the criterion with $\log (\mathrm{N/O}) \gtrsim -1.0$, which are excluded from the $\Yp$ determination. We do not detect [N \textsc{ii}] $\lambda\lambda$6548,6583 in J0228-0210 probably due to the instrumental noise. We thus do not use this galaxy for the $\Yp$ determination.\footnote[4]{Even if we relax the criterion of low N/O, $\Yp$ increases only $\sim0.3\sigma$ level, which does not change our conclusion.}

\subsection{Final Sample}
In Table \ref{table:abundance}, we present whether each galaxy satisfy the three criteria (low N/O, negligible $y^0$ , and reproduction with {\tt\string YMCMC}). We find that 9 galaxies satisfy all of the three criteria. We add these 9 galaxies to the 58 galaxies in the litarature sample (Section \ref{sec:literature_gal}), obtaining 67 galaxies as a final sample for $\Yp$ determination.

\begin{deluxetable*}{ccccccccccc}
\tablewidth{12pt}
\tablecaption{Chemical abundance ratios}
\tablehead{ 
\colhead{ID} & \colhead{$\rm O/\rm H$} & \colhead{$\log \mathrm{(N/O)}$} & \colhead{$\log \eta_\mathrm{rds}$} & \colhead{$y^{+}$} & \colhead{$y^{++}$} & \colhead{$y$} & \multicolumn{3}{c}{Criteria}  & Final sample \\
\cline{8-10}  &  \colhead{($\times10^{-5}$)} & \colhead{} & \colhead{} & \colhead{} & \colhead{} & \colhead{} & N/O & $\eta_\mathrm{rds}$ & YMCMC & \colhead{} 
}
\startdata
J2115-1734  &  $\err{4.39}{0.09}{0.09}$ & $\err{-1.42}{0.03}{0.03}$ & $\err{0.21}{0.01}{0.01}$ & $\err{0.0740}{0.0031}{0.0022}$$^\star$ & $\err{0.0030}{0.0001}{0.0001}$ & $\err{0.0770}{0.0031}{0.0022}$ & Yes & Yes & No & No \\
J0159+0751  &  $\err{3.66}{0.42}{0.42}$ & $\err{-1.33}{0.14}{0.14}$ & $\err{-0.38}{0.43}{0.43}$ & $\err{0.0719}{0.0022}{0.0019}$$^\star$ & $\err{0.0013}{0.0006}{0.0006}$ &  $\err{0.0732}{0.0022}{0.0019}$ & Yes & Yes & No & No \\
J2302+0049  & $\err{4.38}{0.19}{0.19}$ & $\err{-1.74}{0.06}{0.06}$ & $\err{0.01}{0.04}{0.04}$ & $\err{0.0723}{0.0030}{0.0029}$$^\star$ & $\err{0.0022}{0.0003}{0.0003}$ & $\err{0.0745}{0.0030}{0.0029}$ & Yes & Yes & No & No \\
%J2229+2725  &  $\err{1.22}{0.09}{0.09}$ & $\err{-1.54}{0.18}{0.18}$ & $\err{-0.62}{0.19}{0.19}$ & $\err{0.0935}{0.0217}{0.0122}$ & $\err{0.0020}{0.0002}{0.0002}$ & $\err{0.0955}{0.0217}{0.0122}$ & Yes & Yes & Yes & Yes\\
J2229+2725  &  $\err{1.22}{0.09}{0.09}$ & $\err{-1.54}{0.18}{0.18}$ & $\err{-0.62}{0.19}{0.19}$ & $\err{0.0731}{0.0037}{0.0034}$$^\star$ & $\err{0.0020}{0.0002}{0.0002}$ & $\err{0.0751}{0.0037}{0.0034}$ & Yes & Yes & No & No\\
J0036+0052  &  $\err{3.62}{0.21}{0.21}$ & $\err{-1.56}{0.08}{0.06}$ & $\err{0.25}{0.03}{0.04}$ & $\err{0.0896}{0.0079}{0.0065}$ & $\err{0.0011}{0.0003}{0.0003}$ & $\err{0.0907}{0.0079}{0.0065}$ & Yes & Yes & Yes & Yes \\
J0808+1728  &  $\err{3.85}{0.40}{0.40}$ & $\err{-1.44}{0.10}{0.23}$ & $\err{-0.15}{0.07}{0.06}$ & $\err{0.0926}{0.0170}{0.0122}$ & $\cdots$ & $\err{0.0926}{0.0170}{0.0122}$ & Yes & Yes & Yes & Yes \\
J2136+0414  &  $\err{2.39}{0.11}{0.11}$ & $\err{-1.51}{0.08}{0.09}$ & $\err{-0.06}{0.03}{0.03}$ & $\err{0.0925}{0.0051}{0.0040}$$^\star$ & $\err{0.0020}{0.0005}{0.0005}$ & $\err{0.0945}{0.0051}{0.0040}$ & Yes & Yes & No & No \\
J0210-0124  &  $\err{5.72}{0.13}{0.13}$ & $\err{-1.81}{0.02}{0.02}$ & $\err{0.45}{0.02}{0.02}$ & $\err{0.1321}{0.0124}{0.0221}$$^\star$ & $\cdots$ & $\err{0.1321}{0.0124}{0.0221}$ & Yes & Yes & No & No \\
%J2104-0035  &&&&&&&&\\
J0159-0622  &  $\err{2.44}{0.05}{0.05}$ & $\err{-1.14}{0.02}{0.02}$ & $\err{0.29}{0.01}{0.01}$ & $\err{0.0902}{0.0019}{0.0020}$$^\star$ & $\cdots$ & $\err{0.0902}{0.0019}{0.0020}$ & Yes & Yes & No & No \\
%J0107+0103  &&&&&&&&\\
J0134-0038  &  $\err{27.52}{11.70}{11.70}$ & $\err{-2.17}{0.29}{0.27}$ & $\err{0.69}{0.29}{0.29}$$^\ddagger$ & $\err{0.0829}{0.0057}{0.0057}$ & $\cdots$ & $\err{0.0829}{0.0057}{0.0057}$ & Yes & No & Yes & No  \\
J0811+4730  &  $\err{1.13}{0.27}{0.27}$ & $\err{-0.91}{0.20}{0.22}$$^\dagger$ & $\err{-0.05}{0.13}{0.14}$ & $\err{0.0932}{0.0078}{0.0078}$ & $\cdots$ & $\err{0.0932}{0.0078}{0.0078}$ & No & Yes & Yes & No \\
J0845+0131  &  $\err{2.05}{0.20}{0.20}$ & $\err{-0.95}{0.08}{0.07}$$^\dagger$ & $\err{-0.21}{0.04}{0.08}$ & $\err{0.1026}{0.0030}{0.0032}$$^\star$ & $\cdots$ & $\err{0.1026}{0.0030}{0.0032}$ & No & Yes & No & No \\
J2314+0154  &  $\err{1.86}{0.44}{0.44}$ & $\err{-1.57}{0.21}{0.23}$ & $\err{0.34}{0.12}{0.14}$ & $\err{0.0773}{0.0057}{0.0057}$ & $\cdots$ & $\err{0.0773}{0.0057}{0.0057}$ & Yes & Yes & Yes & Yes \\
J0007+0226  &  $\err{9.54}{0.46}{0.46}$ & $\err{-1.34}{0.07}{0.07}$ & $\err{-0.78}{0.03}{0.04}$ & $\err{0.0910}{0.0083}{0.0063}$ & $\err{0.0012}{0.0003}{0.0003}$ & $\err{0.0922}{0.0083}{0.0063}$ & Yes & Yes & Yes & Yes \\
J0014-0043  &  $\err{13.80}{1.22}{1.22}$ & $\err{-1.60}{0.07}{0.07}$ & $\err{0.70}{0.07}{0.08}$ & $\err{0.0864}{0.0071}{0.0043}$ & $\err{0.0015}{0.0003}{0.0003}$ &  $\err{0.0879}{0.0071}{0.0043}$ & Yes & Yes & Yes & Yes \\
J0226-5017  &  $\err{4.48}{0.26}{0.26}$ & $\err{-1.84}{0.08}{0.06}$ & $\err{1.19}{0.06}{0.05}$$^\ddagger$ & $\err{0.1230}{0.0078}{0.0098}$$^\star$ & $\cdots$ & $\err{0.1230}{0.0078}{0.0098}$ & Yes & No & No & No\\
J0228-0210  &  $\err{1.52}{0.14}{0.14}$ & $\cdots^{b}$ & $\err{0.52}{0.06}{0.07}$ & $\err{0.1253}{0.0193}{0.0610}$$^\star$ & $\err{0.0045}{0.0011}{0.0011}$ & $\err{0.1298}{0.0193}{0.0610}$ & No & Yes & No & No \\
J0248-0817  &  $\err{12.05}{0.30}{0.30}$ & $\err{-1.28}{0.03}{0.03}$ & $\err{0.13}{0.04}{0.03}$ & $\err{0.0871}{0.0015}{0.0012}$$^\star$ & $\err{0.0006}{0.0001}{0.0001}$ & $\err{0.0877}{0.0015}{0.0012}$ & Yes & Yes & No & No \\
SBS0335E  &  $\err{2.04}{0.05}{0.05}$ & $\err{-1.52}{0.02}{0.02}$ & $\err{-0.24}{0.04}{0.04}$ & $\err{0.0731}{0.0024}{0.0026}$$^\star$ & $\err{0.0012}{0.0001}{0.0001}$ & $\err{0.0743}{0.0024}{0.0026}$ & Yes & Yes & No & No\\
J0833+2508 &  $\err{2.43}{0.77}{0.77}$ & $\err{-1.33}{0.23}{0.12}$ & $\err{1.03}{0.23}{0.18}$$^\ddagger$ & $\err{0.0929}{0.0123}{0.0113}$$^\star$ & $\cdots$ & $\err{0.0929}{0.0123}{0.0113}$ & Yes & No & No & No\\
Mrk996 &  $\err{1.64}{0.05}{0.06}$ & $\err{-0.78}{0.04}{0.04}$$^\dagger$ & $\err{0.66}{0.15}{0.15}$ & $\err{0.1336}{0.0111}{0.0141}$$^\star$ & $\err{0.0079}{0.0001}{0.0001}$ & $\err{0.1415}{0.0111}{0.0141}$ & No & Yes & No & No \\
J0335-0038 &  $\err{7.03}{0.42}{0.42}$ & $\err{-1.74}{0.08}{0.08}$ & $\err{0.26}{0.05}{0.08}$ & $\err{0.0899}{0.0080}{0.0055}$ & $\cdots$ & $\err{0.0899}{0.0080}{0.0055}$ & Yes & Yes & Yes & Yes \\
J0301+0114 &  $\err{12.46}{1.07}{1.07}$ & $\err{-1.44}{0.09}{0.09}$ & $\err{-0.28}{0.05}{0.05}$ & $\err{0.0980}{0.0087}{0.0066}$ & $\cdots$ & $\err{0.0980}{0.0087}{0.0066}$ & Yes & Yes & Yes & Yes \\
J0313+0006  &  $\err{6.83}{0.44}{0.44}$ & $\err{-1.96}{0.11}{0.13}$ & $\err{-0.42}{0.05}{0.05}$ & $\err{0.0645}{0.0052}{0.0032}$$^\star$ & $\cdots$ & $\err{0.0645}{0.0052}{0.0032}$ & Yes & Yes & No & No \\
J0825+1846 &  $\err{4.84}{0.20}{0.20}$ & $\err{-1.49}{0.06}{0.06}$ & $\err{-0.10}{0.03}{0.03}$ & $\err{0.0737}{0.0031}{0.0029}$$^\star$ & $\err{0.0017}{0.0003}{0.0003}$ & $\err{0.0754}{0.0031}{0.0029}$ & Yes & Yes & No & No\\
J0815+2156  &  $\err{8.87}{0.49}{0.49}$ & $\err{-1.46}{0.06}{0.06}$ & $\err{-0.16}{0.04}{0.04}$ & $\err{0.0871}{0.0055}{0.0047}$ & $\err{0.0011}{0.0003}{0.0003}$ & $\err{0.0882}{0.0055}{0.0047}$  & Yes & Yes & Yes & Yes \\
I Zw 18 NW  & $\err{1.34}{0.15}{0.15}$ & $\err{-1.59}{0.56}{0.56}$ & $\err{-0.14}{0.44}{0.44}$ & $\err{0.0684}{0.0033}{0.0024}$ & $\err{0.0032}{0.0005}{0.0005}$ & $\err{0.0716}{0.0033}{0.0024}$ & Yes & Yes & Yes & Yes\\
\enddata
\tablecomments{
$^\dagger\log\mathrm{(N/O)>-1.0}$ within the $1\sigma$ level. \\
$^\ddagger\log\mathrm{(\eta_{rds})>0.9}$ within the $1\sigma$ level. \\
$^\star$Not all of the detected emission lines are reproduced with {\tt\string YMCMC} within the $2\sigma$ levels. \\
$^b$[N \textsc{ii}] $\lambda\lambda$6548,6583\AA\, lines are not detected due to the nearby artifacts.
} 
\label{table:abundance}
\end{deluxetable*}

\section{Result}\label{sec:result}
Using the He and O abundances of our final sample, we present our determinations of $\Yp$ with several methods and assess possible systematics in Section \ref{sec:Yp}. We then compare our results with the previous $\Yp$ measurements in Section \ref{sec:Yp_comparison}.

\subsection{Primordial Helium Abundance}\label{sec:Yp}
\begin{figure*}
    \centering
    % Keep it two-column wide, but cap the height so it won't become a float-only page.
    \includegraphics[width=0.95\textwidth,height=0.55\textheight,keepaspectratio]{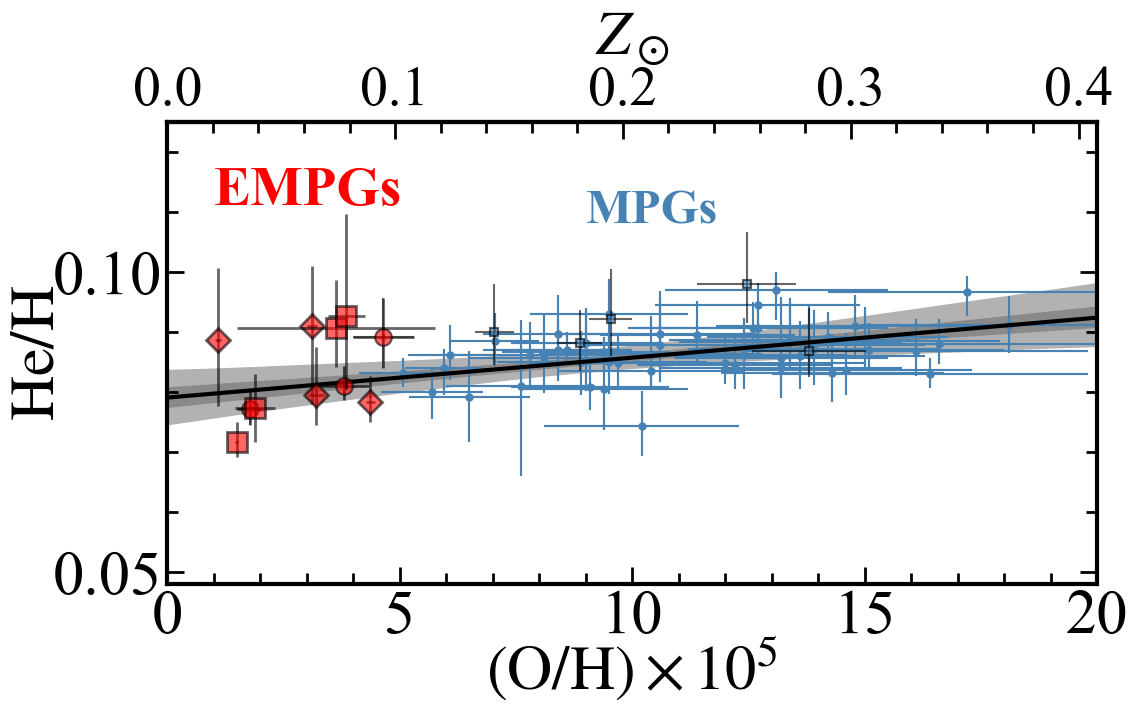}
    \caption{Helium abundance as a function of oxygen abundance.
    The red and blue data points denote the EMPGs and MPGs, respectively. The squares, diamonds, and circles represent the galaxies presented in this study, \cite{Matsumoto+2022}, and \cite{Hsyu+2020}, respectively. The black line shows the best-fit linear model, while the dark and light gray shaded regions are the 1$\sigma$ and 3$\sigma$ errors of the linear fitting, respectively.}
    \label{fig:linearfit}
\end{figure*}

Helium is produced both by BBN and stellar nucleosynthesis, while oxygen is made only by stellar nucleosynthesis. The primordial helium abundance is thus the helium abundance
%by estimating the helium abundance 
at the zero oxygen abundance. 
The standard procedure to derive the primordial helium abundance is to derive a linear correlation on a $y-\mathrm{(O/H)}$ plane:
\begin{equation}
    y = \yp + \frac{\mathrm{d}y}{\mathrm{d(O/H)}}\mathrm{(O/H)},
\end{equation}
where $\yp$ is the primordial number abundance ratio of helium to hydrogen. To determine $\yp$ and $\mathrm{d}y/\mathrm{d(O/H)}$, we maximize the likelihood function given by
\begin{align}\label{eq:log_likeli}
\begin{split}
    \log\left(\mathcal{L}\right) = -\frac{1}{2}\sum_{i}&\left[\frac{\left(y_i - a\mathrm{\left(\frac{O}{H}\right)}_i-b\right)^2}{\sigma_{y_i}^2 + a^2\sigma_{\mathrm{(O/H)}_i}^2+\sigma_\mathrm{int}^2} \right.\\
    & \left.+ \log\left(\sigma_{y_i}^2 + a^2\sigma_{\mathrm{(O/H)}_i}^2+\sigma_\mathrm{int}^2\right)\right],
\end{split}
\end{align}
where $a \equiv \mathrm{d}y/\mathrm{d(O/H)}$ is the slope and $b \equiv \yp$ is the intercept. 
Here, $y_i$ and $(\mathrm{O}/\mathrm{H})_i$ denote the $y$ and O/H values of the galaxy $i$, respectively, while $\sigma_{y_i}$ and $\sigma_{(\mathrm{O}/\mathrm{H})_i}$ are the corresponding $1\sigma$ errors. We introduce the intrinsic scatter $\sigma_\mathrm{int}$ to explain unrecognized scatters of the He/H measurements following the previous primordial abundance studies \citep{Cooke+2018_Dp, Hsyu+2020, Matsumoto+2022}. The summation is over all galaxies in the sample. Using the MCMC algorithm, we obtain our fiducial result of

\begin{align}
\label{eq:yp_result_fiducial}
    \yp &= \err{0.0790}{0.0017}{0.0017}\\
    \frac{\mathrm{d}y}{\mathrm{d(O/H)}} &= \err{0.00067}{0.00017}{0.00016}\\
    \sigma_{\mathrm{int}} &= \err{0.0010}{0.0010}{0.0007}.
\end{align}
In Figure \ref{fig:linearfit}, we plot our final sample galaxies on the $y-\mathrm{(O/H)}$ plane, together with the best-fit linear model. We find that increasing the number of EMPGs introduces less extrapolation into the result than the previous work. %significantly reduce the extrapolation distance.

The discussions of the primordial helium abundance have been made with the helium mass abundance, $\Yp$, instead of the helium number abundance, $\yp$. To compare our measurement to the previous results, we convert our $\yp$ value to the $\Yp$ value using the relation

\begin{equation}\label{eq:yp_Yp}
    Y_{\mathrm{P}} = \frac{4y_\mathrm{P}}{1+4y_\mathrm{P}}.
\end{equation}
Equation \eqref{eq:yp_result_fiducial} thus converts to

\begin{equation}\label{eq:Yp_fiducial}
    \Yp = \err{0.2402}{0.0040}{0.0040}.
\end{equation}

\cite{Peimbert_2007} have discussed the largest systematics in $\Yp$ determination is the collisional excitation of H \textsc{i} lines, which is estimated to be $\Delta \Yp = \pm0.0015$. To assess possible systematics arising from the different atomic data, we conduct {\tt\string YMCMC} analysis using different collision strength data from the default one \citep{Anderson+2002}. We use collision strength from CHIANTI atomic database \citep{Dere+1997, Young+2016}. In this case, we obtain

\begin{align}\label{eq:yp_result_chianti}
    \yp&=\err{0.0783}{0.0016}{0.0013}\\
    \frac{\mathrm{d}y}{\mathrm{d(O/H)}} &= \err{0.00071}{0.00014}{0.00015}\\
    \sigma_\mathrm{int} &= \err{0.0010}{0.0009}{0.0007}.
\end{align}
This $\yp$ value yields

\begin{equation}
    \Yp = \err{0.2385}{0.0037}{0.0031}.
\end{equation}
We find that this $\Yp$ value is lower than the fiducial result of $\Yp=\err{0.2402}{0.0040}{0.0040}$ (Equation \eqref{eq:Yp_fiducial}) by 0.0017, which is comparable to the systematic uncertainty estimated in \cite{Peimbert_2007}, and is not dominant compared to the current statistical uncertainty.

\begin{figure*}
    \centering
    \includegraphics[width=1\linewidth]{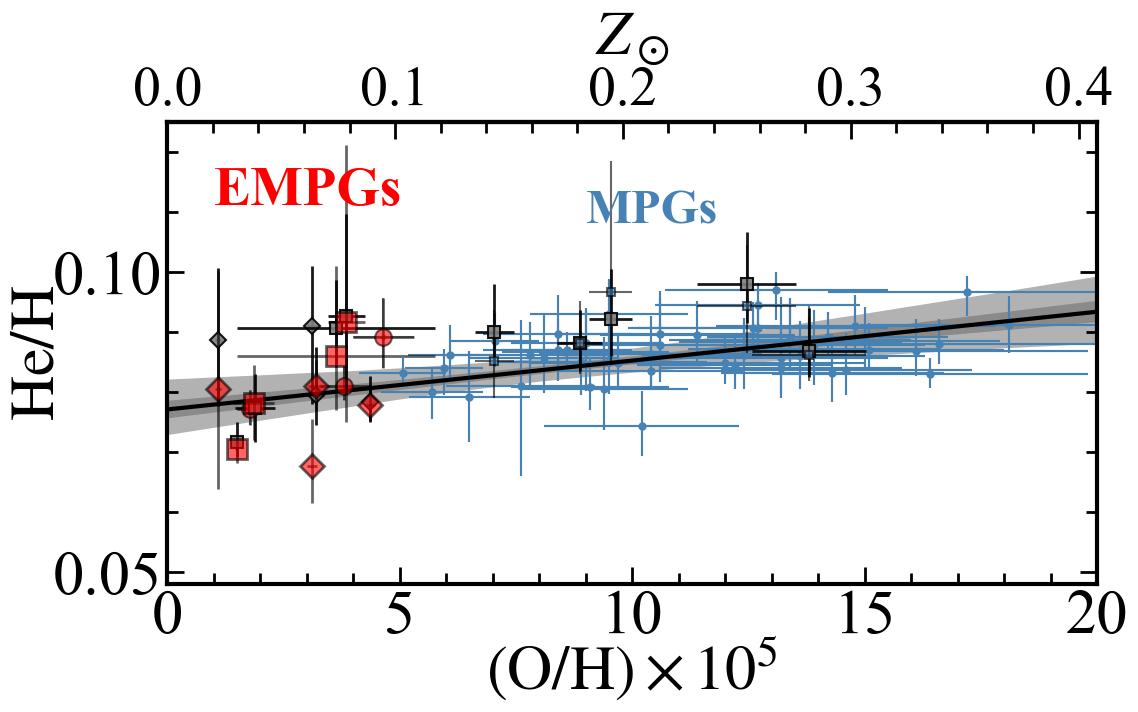}
    \caption{Same as Figure \ref{fig:linearfit}, but the $y^+$ values of the galaxies presented in this study are derived with the prior distributions that allows negative values for $c(\mathrm{H}\beta),~a_\mathrm{H},~ a_\mathrm{He}$, and $\tau_\mathrm{He}$. The gray squares are the values derived with the prior distributions that do not allow these four parameters to be negative (i.e., the values shown in Figure \ref{fig:linearfit}). Note that the gray data points are not used in the linear regression.  %except for the gray squares, which represent the values derived with the prior distributions excluding the unphysical values (Equation \eqref{eq:prior}). Note that the gray data points are not used in the linear regression.
    }
    \label{fig:linearfit_negativeprior}
\end{figure*}

\citet{Matsumoto+2022} allow $c(\mathrm{H}\beta)$, $a_\mathrm{H}$, $a_\mathrm{He}$, and $\tau_\mathrm{He}$ to take negative values to avoid systematic errors. \citet{Matsumoto+2022} claim that the observed flux ratios are scattered around the ``true" flux ratios by statistical uncertainties,
and that the best-fit parameters are also scattered around the ``true" parameter values reproducing the ``true" flux ratios. For example, the physical parameters of $c(\mathrm{H}\beta),~a_\mathrm{H},~ a_\mathrm{He}$, and $\tau_\mathrm{He}$ should have positive values. However, if the "true" values of these parameters are positive but close to zero, their best-fit values can be negative by statistical uncertainties. If the prior distributions are limited to the positive values, the best-fit parameters as determined by the median values can be biased to a large positive value significantly beyond the statistical uncertainties. To assess the impact of the choice of the priors, we also conduct {\tt\string YMCMC} fitting allowing the negative values for these parameters and derive $\Yp$. The He and O abundances are shown in Figure \ref{fig:linearfit_negativeprior}.
In this case, we find

\begin{align}\label{eq:yp_result_negativeprior}
    \yp&=\err{0.0770}{0.0015}{0.0015}\\
    \frac{\mathrm{d}y}{\mathrm{d(O/H)}} &= \err{0.00082}{0.00015}{0.00015}\\
    \sigma_\mathrm{int} &= \err{0.0015}{0.0009}{0.0009}.
\end{align}
This $\yp$ value converts to

\begin{equation}\label{eq:Yp_negativeprior}
    \Yp = \err{0.2355}{0.0034}{0.0035}.
\end{equation}
The negative prior distributions yield lower $\Yp$ value probably because of the $y^+$--$a_\mathrm{He}$ correlation (Section \ref{sec:y+}). Hereafter we use $y^+$ values derived with the prior distributions that exclude negative values for $c(\mathrm{H}\beta),~a_\mathrm{H},~ a_\mathrm{He}$, and $\tau_\mathrm{He}$ (Equation \eqref{eq:prior}).

Although we derive best-fit parameters based on the median posterior values following \cite{Hsyu+2020}, \cite{Aver_2015} derive best-fit parameters that maximize likelihood. We summarize the maximum likelihood estimates in Table \ref{tab:mcmc_recovered_params_ml}. Using these parameters, we obtain

\begin{align}\label{eq:yp_result_ml}
    \yp &= \err{0.0791}{0.0017}{0.0015}\\
    \frac{\mathrm{d}y}{\mathrm{d(O/H)}} &= \err{0.00064}{0.00014}{0.00015}\\
    \sigma_\mathrm{int} &= \err{0.0010}{0.0010}{0.0007},
\end{align}
We thus obtain

\begin{equation}
    \Yp = \err{0.2404}{0.0039}{0.0034},
\end{equation}
which is in good agreement with our fiducial result.

\begin{figure}
    \centering
    \includegraphics[width=1\linewidth]{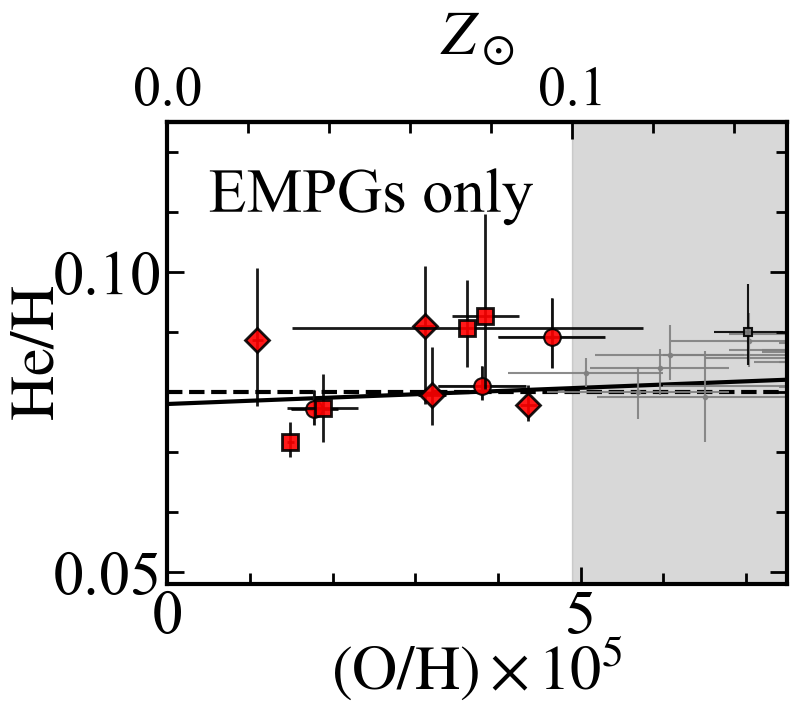}
    \caption{Same as Figure \ref{fig:linearfit}, but for the results derived only with EMPGs. The symbols are the same as Figure \ref{fig:linearfit} except for the MPGs, which are shown in the gray points. The gray shaded region represents the metallicity region corresponding to the MPGs ($\gtrsim 0.1\,Z_\odot$).}
    \label{fig:linearfit2}
\end{figure}

To investigate the possible systematics arising from the chemical evolution, 
we also derive $\Yp$ with EMPGs alone (Figure \ref{fig:linearfit2}). We first assume the slope $\mathrm{d}y/\mathrm{d(O/H)}$ value to be $\mathrm{d}y/\mathrm{d(O/H)}=0.00054$, which is the one derived by \cite{Hsyu+2020}. We obtain

\begin{align}\label{eq:yp_result3}
    \yp&=\err{0.0782}{0.0025}{0.0018}\\
    \sigma_\mathrm{int} &= \err{0.0035}{0.0030}{0.0022}.
\end{align}
The $\sigma_\mathrm{int}$ value is larger than the one derived with both EMPGs and MPGs (Equations \eqref{eq:yp_result_fiducial}, \eqref{eq:yp_result_chianti}, and \eqref{eq:yp_result_negativeprior}), which implies that EMPGs have larger intrinsic scatter of He/H than MPGs. Equation \eqref{eq:yp_Yp} yields

\begin{equation}\label{eq:Yp_EMPG_only1}
    \Yp = \err{0.2383}{0.0058}{0.0043}.
\end{equation}
However, it is uncertain whether the slope in the EMPG regime is similar to that of the MPG regime. The result of Equation \eqref{eq:Yp_EMPG_only1} might thus contain potential bias arising from the assumption of the slope. To be conservative, we fix the slope to be zero, i.e., assuming no He/H evolution. In this case, we obtain

\begin{align}\label{eq:yp_result4}
    \yp&=\err{0.0799}{0.0027}{0.0020}\\
    \sigma_\mathrm{int} &= \err{0.0043}{0.0031}{0.0024}.
\end{align}
Using Equation \eqref{eq:yp_Yp}, we obtain

\begin{equation}
    \Yp = \err{0.2422}{0.0060}{0.0042}.
\end{equation}

\begin{figure*}[tb]
    \includegraphics[width=1\textwidth]{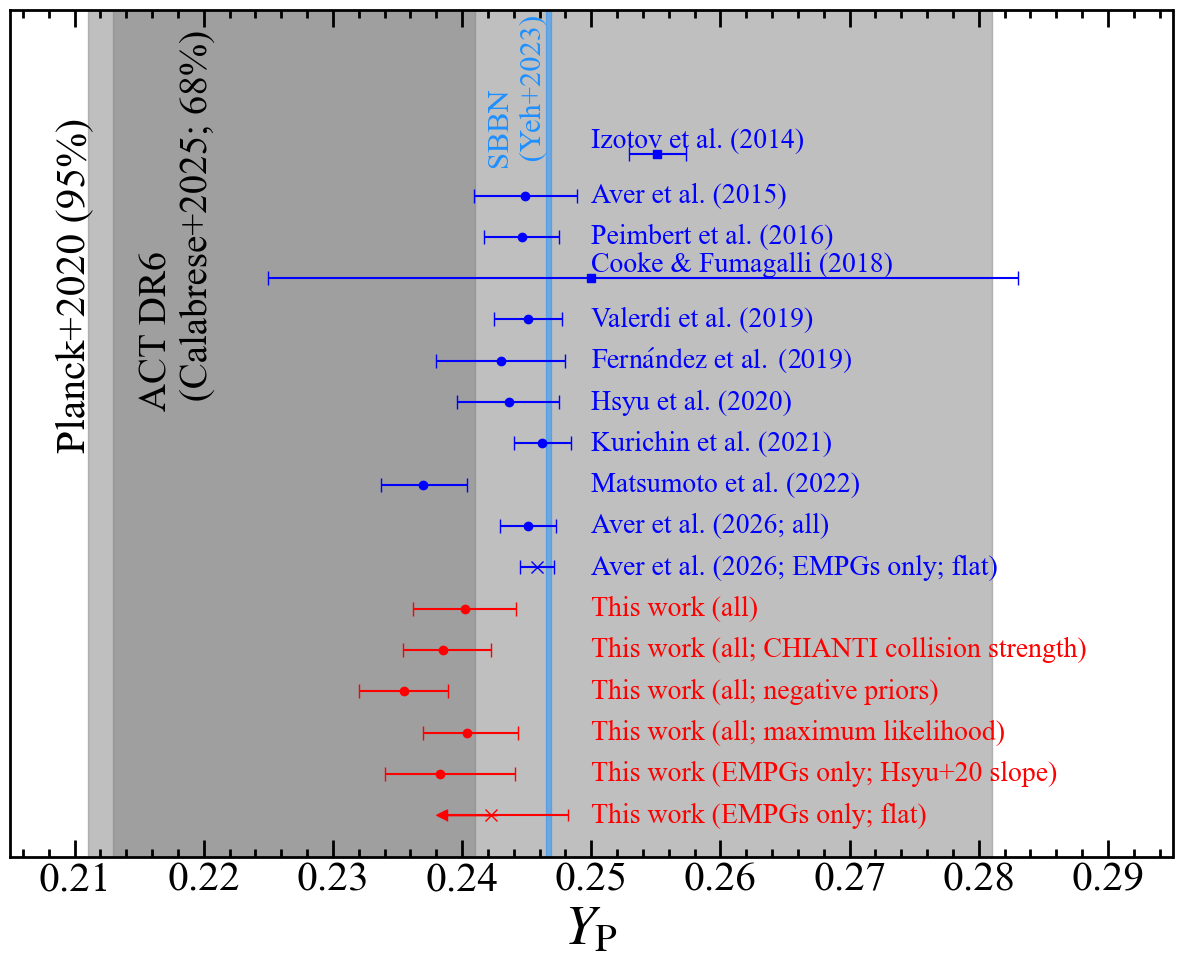}
    \centering
    \caption{Comparison of our $\Yp$ values with those derived in previous studies. The red circles show our $\Yp$ values. The blue circles represent the previous $\Yp$ values derived from the observations of the metal-poor H \textsc{ii} regions (\citealt{Peimbert+2016, Velardi+2019, Fernandez+2019, Hsyu+2020, Kurichin+2021, Matsumoto+2022}, \citealt{Aver+2026}) and the quasar absorption system \citep{Cooke+2018}. The crosses indicate values derived from zero-slope fits using only EMPGs. All error bars denote the 1$\sigma$ uncertainties. The light gray shaded region presents the constraint on $\Yp$ from the CMB and BAO observations with the 95\% confidence level \citep{Planck2020}, while the dark gray shaded region indicates that with the 68\% confidence level from ACT \citep{Calabrese+2025}. The blue shaded region indicates the standard BBN prediction from \cite{Yeh+2023}.}
    \label{fig:Yp_comparison}
\end{figure*}

\begin{table}[tb]
\centering
\caption{Comparison of $Y_\mathrm{P}$ values}
\label{tab:Yp_comparison}
\begin{tabular}{lc}
\hline
Reference & $Y_\mathrm{P}$\\
\hline
\cite{Izotov+2014} & $0.2551^{+0.0022}_{-0.0022}$\\
\cite{Aver_2015} & $0.2449^{+0.0040}_{-0.0040}$\\
\cite{Peimbert+2016} & $0.2446^{+0.0029}_{-0.0029}$\\
\cite{Cooke+2018} & $0.250^{+0.033}_{-0.025}$\\
\cite{Velardi+2019} & $0.2451^{+0.0026}_{-0.0026}$\\
\cite{Fernandez+2019} & $0.243^{+0.005}_{-0.005}$\\
\cite{Hsyu+2020} & $0.2436^{+0.0039}_{-0.0040}$\\
\cite{Planck2020} & $0.247^{+0.0034}_{-0.0036}$$^\dagger$\\
\cite{Kurichin+2021} & $0.2462^{+0.0022}_{-0.0022}$\\
\cite{Matsumoto+2022} & $0.2370^{+0.0034}_{-0.0033}$\\
\cite{Yeh+2023} & $0.2467^{+0.0002}_{-0.0002}$ \\
\cite{Calabrese+2025} & $0.227^{+0.14}_{-0.14}$ \\
\cite{Aver+2026} & $0.2458^{+0.0013}_{-0.0013}$\\
This work (all; fiducial) & $0.2402^{+0.0040}_{-0.0040}$ \\
This work (all; CHIANTI collision strength) & $0.2385^{+0.0037}_{-0.0031}$\\
This work (all; negative priors) & $0.2355^{+0.0034}_{-0.0035}$\\
This work (all; maximum likelihood) & $0.2404^{+0.0039}_{-0.0034}$\\
This work (EMPGs only; Hsyu+20 slope)  & $0.2383^{+0.0058}_{-0.0043}$\\
This work (EMPGs only; flat) & $0.2422^{+0.0060}_{-0.0042}$\\
\hline
\end{tabular}

{\footnotesize\raggedright$^\dagger$ Error is given by 95\% confidence range, while the others are given by 68\%.\par}
\end{table}

\subsection{Comparison with the Previous Measurements}\label{sec:Yp_comparison}
We compare these results with the previous studies in Table \ref{tab:Yp_comparison} and Figure \ref{fig:Yp_comparison}. The $\Yp$ measurements of our study are comparable with the previous measurement of \cite{Matsumoto+2022}. Interestingly, our results are also consistent with the recent CMB constraints with ACT, which shows slightly low $\Yp$ values \citep{Calabrese+2025}. 
However, our measurements are lower than the standard BBN (SBBN) prediction \citep{Yeh+2023} at more than 1$\sigma$ level, and also lower than the other observational constraints at the $\sim1\sigma$ level, except for the most conservative one (``EMPGs only; flat" in Figure \ref{fig:Yp_comparison}).  %The main difference between our result from most of these studies is the increased number of EMPGs, which were limited in previous works (three EMPGs in \citealt{Hsyu+2020}). Although the uncertainties are comparable to the previous studies probably because of the large number of MPGs that is originally included, the increase in the extremely metal-poor regime leads to the slightly lower $\Yp$ values. One exception is the most recent work by \cite{Aver+2026}, who have used 19 EMPGs. The difference between our result and \cite{Aver+2026} is the slope of the linear regression. \cite{Aver+2026} have derived the slope value $\mathrm{d}y/\mathrm{d(O/H)}=\err{0.00014}{0.00012}{0.00012}$, which is $\sim5$ times smaller than that in our fiducial measurement ($\mathrm{d}y/\mathrm{d(O/H)}=\err{0.00072}{0.00014}{0.00015}$; Equation \eqref{eq:yp_result_fiducial}). This difference may cause the difference in the $\Yp$ value. Nevertheless, \cite{Aver+2026} measurement with zero-slope fitting to EMPGs is in good agreement with that of our measurement. These comparison demonstrate that the slope value also have significant effect on the $\Yp$ measurement.}
The main difference between our result and most previous studies is the increased number of EMPGs (e.g., only three EMPGs in \citealt{Hsyu+2020}). Although the overall uncertainties are comparable to those in previous studies--likely reflecting the large number of MPGs included--the expanded sampling in the extremely metal-poor regime may contribute to slightly lower inferred $\Yp$ values.
An exception is the recent work by \cite{Aver+2026}, who analyzed 19 EMPGs and 22 MPGs. A key difference between our result and theirs lies in the slope of the linear regression. \cite{Aver+2026} derived a slope of $\mathrm{d}y/\mathrm{d(O/H)}=\err{0.00014}{0.00012}{0.00012}$, which is approximately five times smaller than our fiducial value ($\mathrm{d}y/\mathrm{d(O/H)}=\err{0.00068}{0.00016}{0.00016}$; Equation~\eqref{eq:yp_result_fiducial}), probably because of the smaller number of MPGs in \cite{Aver+2026}. This difference likely contributes to the offset in the inferred $\Yp$ values. Nevertheless, their zero-slope fit restricted to EMPGs agrees with our corresponding measurement (Equation \eqref{eq:yp_result4}; the lowest data point in Figure \ref{fig:Yp_comparison}) within the uncertainties. These comparisons demonstrate that the adopted slope has a significant impact on the derived $\Yp$ value.

%Interestingly, except for the measurements of \cite{Izotov+2014} and \cite{Cooke+2018}, all of the measurements are smaller than the SBBN prediction. If the SBBN prediction represents the true $\Yp$ value, the measurements should scatter around it. This implies that the true $\Yp$ value is slightly smaller than that of the SBBN prediction, or that there are systematic uncertainties in the $\Yp$ measurements that we have not recognized.

\section{Discussion}\label{sec:discussion}
\begin{figure}[tb]
    \includegraphics[width=0.45\textwidth]{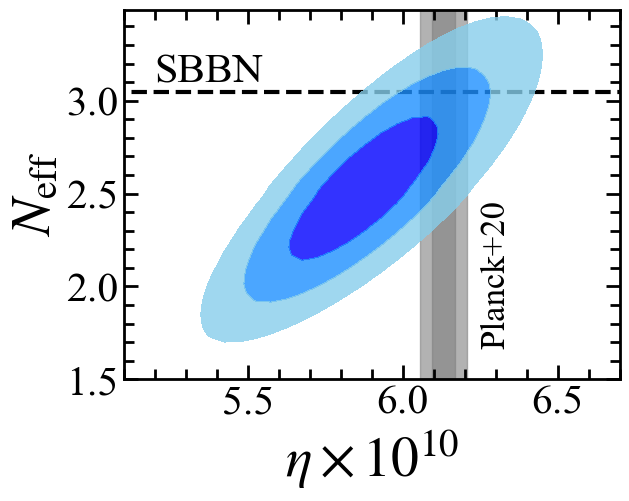}
    \caption{Observational constraints on $\neff$ and $\eta$. The blue contours present the 1$\sigma$, 2$\sigma$, and 3$\sigma$ confidence levels of this work. The gray contours show the 1$\sigma$ and 2$\sigma$ confidence levels from the CMB measurement by \cite{Planck2020}. The dashed line indicates the standard BBN value of $\neff=3.044$.}
    \label{fig:Neff_eta}
\end{figure}

The primordial helium and deuterium abundances depend on the cosmological parameters such as the baryon-to-photon ratio, $\eta$, and $N_\mathrm{eff}$. We place constraints on these parameters with our $\Yp$ measurement and the recent primordial deuterium abundance measurement of $\mathrm{(D/H)_P} = (2.527 \pm 0.030) \times 10^{-5}$ \citep{Cooke+2018_Dp}. We minimize

\begin{multline}\label{eq:Neff_eta_chi2}
    \chi^2(\eta, N_\mathrm{eff}) = \frac{\left(Y_\mathrm{P, obs}-Y_{\mathrm{P, mod}}(\eta, N_\mathrm{eff})\right)^2}{\sigma^2_{Y_\mathrm{P,obs}}+\sigma^2_{Y_\mathrm{P,mod}}}\\
    +\frac{\left[\mathrm{(D/H)_{P, obs}}-\mathrm{(D/H)_{P, mod}}(\eta, N_\mathrm{eff})\right]^2}{\sigma^2_\mathrm{(D/H)_{P,obs}} + \sigma^2_\mathrm{(D/H)_{P,mod}}},
\end{multline}
where the subscripts `obs' and `mod' indicate the values from observations and models, respectively, and $\sigma$ denotes the corresponding error. We calculate $Y_{\mathrm{P, mod}}(\eta, N_\mathrm{eff})$ and $\mathrm{(D/H)_{P, mod}}(\eta, N_\mathrm{eff})$ with {\tt\string PArthENoPE} (version 3.0; \citealt{Gariazzo+2022}) by varying $\eta$ and $N_\mathrm{eff}$. We use the neutron lifetime $\tau_\mathrm{n} = 879.4 \pm 0.6 \, \mathrm{s}$ \citep{ PDG2020} and the nuclear reaction rates taken from \cite{Pisanti_2021}. 
We adopt
$\sigma^2_\mathrm{(D/H)_{P,mod}} = (0.06)^2 \times 10^{-10}$
corresponding to the uncertainties of the nuclear reaction rates.
%, which is due to the uncertainty of the nuclear reaction rates. 
We use $\sigma^2_{Y_{\mathrm{P,mod}}} = (0.00003)^2 + (0.00012)^2$, where the two terms 
indicate
%correspond to 
the uncertainties of the nuclear reaction rates and the neutron lifetime, respectively. 
We obtain

\begin{align}\label{eq:Neff_eta_result}
    N_\mathrm{eff} &= \err{2.54}{0.20}{0.25} \\
    \eta\times10^{10} &= \err{5.88}{0.15}{0.10}.
\end{align}
Figure \ref{fig:Neff_eta} shows our constraints on $\neff$ and $\eta$. Our slightly low $\Yp$ value results in the mild tension ($\Delta \chi^2 = 6.1$) with the standard BBN value of $\neff = 3.044$ and $\eta$ constraint of Planck. 

%To explain these tensions, in an extension beyond the Standard Model, we allow an asymmetry between electron neutrino, $\nu_\mathrm{e}$, and anti-electron neutrino, $\bar{\nu}_\mathrm{e}$.
To explain these tensions, we consider an extension beyond the Standard Model that allows an asymmetry between electron neutrinos ($\nu_\mathrm{e}$) and antineutrinos ($\bar{\nu}_\mathrm{e}$).
The $\nu_\mathrm{e}-\bar{\nu}_{\mathrm{e}}$ asymmetry (hereafter the lepton asymmetry) shifts the beta equilibrium given by 
\begin{math}
    \mathrm{p} + \mathrm{e}^{-} \leftrightarrow \mathrm{n} + \nu_\mathrm{e}
\end{math},
which changes the neutron abundance before BBN. Because the neutron abundance determines $\Yp$, the lepton asymmetry changes the $\Yp$ value. The lepton asymmetry is represented by the electron neutrino degeneracy parameter, $\xie\equiv \mu_{\nu_\mathrm{e}}/T_{\nu_\mathrm{e}}$ in natural units, where $\mu_{\nu_\mathrm{e}}$ and $T_{\nu_\mathrm{e}}$ are the chemical potential and temperature of electron neutrino, respectively. Using $\xie$, the lepton asymmetry is given by $n_{\nu_\mathrm{e}} - n_{\bar{\nu}_\mathrm{e}} \propto (\pi^2 \xie + \xie^3) T_{\nu_\mathrm{e}}$, where $n_{\nu_\mathrm{e}}$($n_{\bar{\nu}_\mathrm{e}}$) represents the number density of (anti-)electron neutrino (i.e., non-zero $\xie$ leads to the lepton asymmetry). 

To constrain $\xie$ as well as $\neff$ and $\eta$, we minimize
\begin{multline}\label{eq:chi_3}
    \chi^2(\eta, N_\mathrm{eff}, \xie) = \frac{\left(Y_\mathrm{P, obs}-Y_{\mathrm{P, mod}}(\eta, N_\mathrm{eff}, \xie)\right)^2}{\sigma^2_{Y_\mathrm{P,obs}}+\sigma^2_{Y_\mathrm{P,mod}}}\\
    +\frac{\left[\mathrm{(D/H)_{P, obs}}-\mathrm{(D/H)_{P, mod}}(\eta, N_\mathrm{eff}, \xie)\right]^2}{\sigma^2_\mathrm{(D/H)_{P,obs}} + \sigma^2_\mathrm{(D/H)_{P,mod}}}
    +\frac{(\eta-6.132)^2}{0.038^2}.
\end{multline}
Using {\tt\string PArthENoPE}, we calculate $Y_{\mathrm{P, mod}}(\eta, N_\mathrm{eff}, \xie)$ and $\mathrm{(D/H)_{P, mod}}(\eta, N_\mathrm{eff}, \xie)$ by varying $\eta$, $\neff$, and $\xie$. 
To break the degeneracy between these three parameters, we introduce a Gaussian prior, $\eta = (6.132\pm 0.038) \times 10^{-10}$ that is obtained by the CMB measurement with Planck \citep{Planck2020}. 
We find

\begin{align}\label{eq:Neff_eta_xie_result}
 N_\mathrm{eff} &= 3.23^{+0.20}_{-0.26} \\
 \eta\times10^{10} &= 6.14^{+0.03}_{-0.02} \\
 \xie &= 0.05^{+0.02}_{-0.03}.
\end{align}

\begin{figure}[tb]
    \includegraphics[width=0.45\textwidth]{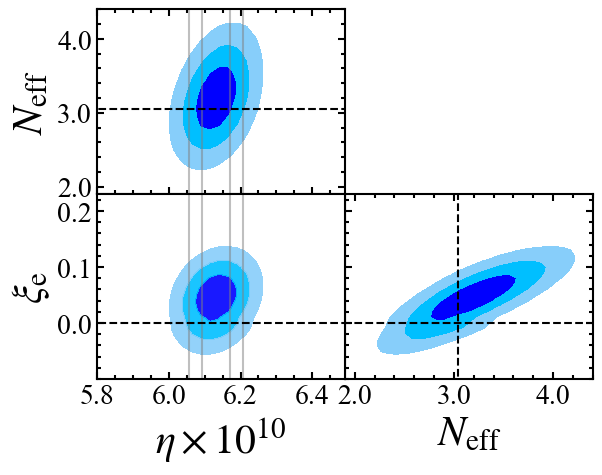}
    \caption{Two dimensional constraints on $\neff$, $\eta$, and $\xie$. The blue contours present the 1$\sigma$, 2$\sigma$, and 3$\sigma$ confidence levels of this work. The black dashed lines denote the standard cosmology values of $\neff = 3.044$ and $\xie = 0$. The gray solid lines show the 1$\sigma$ and 2$\sigma$ confidence levels on $\eta$ from the CMB observation \citep{Planck2020}.}
    \label{fig:Neff_eta_xie}
\end{figure}

In Figure \ref{fig:Neff_eta_xie}, we present our constraints on $\neff$, $\eta$, and $\xie$. Our results indicate the lepton asymmetry at the $\sim 2\sigma$ level.\footnote[5]{We note that if we use the $\Yp$ value obtained with the prior distributions allowing the negative values (Equation \eqref{eq:Yp_negativeprior}), we obtain $\neff=3.30^{+0.13}_{-0.27}, \eta\times10^{10}=6.14^{+0.02}_{-0.03}, \xie=0.07^{+0.02}_{-0.02}$. We also derive $\xie$ fixing $\neff=3.04$ and $\eta\times10^{10}=6.13$, obtaining $\xie=\err{0.03}{0.02}{0.01}$. These results do not change (or even strengthen) our conclusion. }
Although such a large amount of the lepton asymmetry is generally difficult to produce because the lepton number is converted to the baryon number via the sphaleron process, some theoretical studies suggest new mechanisms that %keep the lepton number large 
generate large lepton asymmetry while keeping baryon asymmetry small (e.g., \citealt{Kawasaki_Murai_2022}).

The existence of the lepton asymmetry allows the $\neff$ value comparable to, or even larger than the standard value of $\neff=3.044$, which could alleviate the Hubble tension \citep{Seto&Toda_21}. However, the accuracy of the measurement is still insufficient to conclude the existence of the lepton asymmetry. As a next step, we suggest further increasing the number of EMPGs and deriving $\Yp$ without MPGs to reduce systematics arising from the $y$-O/H slope \citep{Peimbert_2007}, as well as to improve statistics. Although the systematics arising from the atomic data is not dominant compared to the present statistical uncertainties, it is also necessary to further assess the impact of uncertainties in atomic data (especially the collision strengths) on the $\Yp$ determination.

\section{Summary}\label{sec:summary}
We conducted NIR spectroscopic observations with Subaru/SWIMS and MOIRCS for 29 galaxies. We apply the photoionization model to these galaxies to estimate the He abundance of each galaxy. We also derive the oxygen abundance of each EMPG from the optical \textsc{[O iii]}, \textsc{[O ii]}, and \textsc{[S ii]} emission lines. Adding the galaxies from the previous studies, we determine $\Yp$ with a final sample of 67 galaxies. Combining our $\Yp$ result with the previous $\mathrm{(D/H)_P}$ measurement \citep{Cooke+2018_Dp}, we put constraints on $\neff$, $\eta$, and $\xie$. Our main results are summarized below.
\begin{enumerate}
    \item Using the linear correlation between the helium and oxygen abundances, we obtain $\Yp = \err{0.2402}{0.0040}{0.0040}$. Our $\Yp$ result is comparable with the measurement of \cite{Matsumoto+2022} and recent CMB measurement with ACT \citep{Calabrese+2025}. However, our $\Yp$ value is lower than the other $\Yp$ measurements at the $\sim 1\sigma$ level. We assess the possible uncertainties arising from the choice of the prior distributions, atomic data, and chemical evolution, which does not significantly change our conclusion.

    \item Combining our slightly low $\Yp$ measurement with the previous $\mathrm{(D/H)_P}$ result of $\mathrm{(D/H)_P} = (2.527 \pm 0.030) \times 10^{-5}$ \citep{Cooke+2018_Dp}, we obtain $\neff = \err{2.54}{0.20}{0.25}$ and $\eta\times10^{10} = \err{5.88}{0.15}{0.10}$, which result in the tension with the standard BBN prediction of $\neff=3.044$ and the Planck result of $\eta\times10^{10}=\err{6.132}{0.038}{0.038}$, respectively. To mitigate this tension, we allow the variation of the lepton asymmetry of the electron neutrino. We obtain $\neff = \err{3.23}{0.20}{0.26}$, $\eta \times 10^{10} = \err{6.14}{0.03}{0.02}$, and $\xie = \err{0.05}{0.02}{0.03}$, suggesting lepton asymmetry at the $\sim2\sigma$ level. However, further improvement in statistics, especially in the extremely metal-poor regime, is necessary to conclude. 
\end{enumerate}

\section*{Acknowledgments}
We are grateful to Takashi Moriya, Hiroya Umeda, and Takahiro Morishita for the valuable discussions. This work is based on observations made by the SDSS and with the Keck and Magellan Telescopes. This research is based on data collected with the Subaru Telescope operated by the National Astronomical Observatory of Japan. We are honored and deeply grateful for the opportunity to observe the universe from Maunakea, a site of profound cultural, historical, and natural significance in Hawai'i.
This publication is based on work supported by the World Premier International Research Center Initiative (WPI Initiative), MEXT, Japan, KAKENHI 20H00180, 21H04467 (M. Ouchi), 25K07297 (M. Kawasaki), 20K22373, 24K07102 (K. Nakajima), 23KF0289, 24H01825, 24K07027 (K. Kohri), 21J20785 (Y. Isobe), 21K13953 24H00245 (Y. Harikane), 25H00664, 25K01046, 26K07153, 26K00744 (Y. Hirai), 25K07361 (M. Onodera), 21H04489 (H. Yajima), and Core-to-Core Program JSCCA20210003 (H. Yanagisawa) through the Japan Society for the Promotion of Science. H. Yanagisawa was supported by a grant from the Hayakawa Satio Fund awarded by the Astronomical Society of Japan. 
H. Yajima is supported by JST FOREST Program JP-MJFR202Z. 
%J.H. Kim acknowledges the support from the National Research Foundation of Korea (NRF) grants, No. 2021M3F7A1084525 and No. 2020R1A2C3011091, and the Institute of Information \& Communications Technology Planning \& Evaluation (IITP) grant, No. RS-2021-II212068 funded by the Korean government (MSIT).
J.H.K. acknowledges the support from the National Research Foundation of Korea (NRF) grants, Nos. RS-2026-25487912 and RS-2026-25490019, funded by the Korean government (MSIT).

This work was supported by the joint research program of the Institute for Cosmic Ray Research (ICRR), University of Tokyo. 
%\end{acknowledgments}

\appendix
\section{Maximum Likelihood Estimates}
Although \cite{Hsyu+2020} and \cite{Matsumoto+2022} have adopted median values of posterior distributions as their best-fit parameters, it is important to see maximum likelihood estimated values as done in e.g., \cite{Aver_2015, Aver+2026}. Table \ref{tab:mcmc_recovered_params_ml} shows best-fit values based on the maximum likelihood. Based on the minimum $\chi^2$ values, we also examine whether the 2$\sigma$ line-reproduction criterion in {\tt\string YMCMC} (Section \ref{sec:y+}) is consistent with the $\chi^2$-based qualification adopted in \cite{Aver_2015, Aver+2026}. Among the 11 objects that satisfied the {\tt\string YMCMC} qualification, the $\chi^2$ test could be applied to seven objects, as the other four (J2314+0154, J0134-0038, J0335-0038, and J0301+0114) had zero degrees of freedom. Five of the seven objects were consistent with the 95\% confidence interval of the $\chi^2$ distribution, with only J2302+0049 and J0811+4730 lying outside it. Overall, this suggests no strong inconsistency between the {\tt\string YMCMC} qualification and the $\chi^2$-based assessment.

\begin{deluxetable*}{cccccccccc}
\tablewidth{12pt}
\tablecaption{Best recovered parameters from MCMC analysis based on maximum likelihood}
\label{tab:mcmc_recovered_params_ml}
\tablehead{ 
\colhead{ID} & \colhead{$y^{+}$} & \colhead{$T_{\rm e}$} & \colhead{log$_{10}(n_{\rm e}/\rm cm^{-3}$)} & \colhead{$c$(H$\beta$)} & \colhead{$a_{\rm H}$} & \colhead{$a_{\rm He}$} & \colhead{$\tau_{\rm He}$} & \colhead{log$_{10}(\xi$)} & \colhead{$\chi^2$} \\
\colhead{} & \colhead{} & \colhead{[K]} & \colhead{} & \colhead{} & \colhead{[\AA]} & \colhead{[\AA]} & \colhead{} & \colhead{} & \colhead{}
}
\startdata
J2115-1734 & $\err{0.0718}{0.0028}{0.0023}$ & $\err{19920}{1433}{1144}$ & $\err{2.80}{0.09}{0.10}$ & $\err{0.28}{0.01}{0.01}$ & $\err{0.00}{0.13}{0.00}$ & $\err{0.06}{0.15}{0.06}$ & $\err{0.01}{0.21}{0.01}$ & $\err{-5.51}{0.81}{0.49}$ & 750 \\
J0159+0751 & $\err{0.1499}{0.0001}{0.0026}$ & $\err{19850}{1181}{1827}$ & $\err{2.63}{0.10}{0.10}$ & $\err{0.07}{0.02}{0.02}$ & $\err{0.06}{0.25}{0.06}$ & $\err{0.85}{0.45}{0.41}$ & $\err{4.93}{0.07}{0.27}$ & $\err{-5.38}{0.85}{0.62}$ & 84 \\
J2302+0049 & $\err{0.0696}{0.0030}{0.0028}$ & $\err{10160}{1732}{1415}$ & $\err{0.12}{0.41}{0.12}$ & $\err{0.14}{0.03}{0.03}$ & $\err{0.14}{0.51}{0.14}$ & $\err{0.00}{0.18}{0.00}$ & $\err{1.85}{1.08}{0.93}$ & $\err{-2.72}{1.71}{1.82}$ & 17 \\
J2229+2725 & $\err{0.0694}{0.0037}{0.0034}$ & $\err{19950}{473}{975}$ & $\err{2.41}{0.12}{0.13}$ & $\err{0.00}{0.01}{0.00}$ & $\err{0.01}{2.19}{0.01}$ & $\err{0.19}{0.93}{0.19}$ & $\err{4.01}{0.48}{0.62}$ & $\err{-5.64}{0.86}{0.36}$ & 28 \\
J0036+0052 & $\err{0.0860}{0.0079}{0.0065}$ & $\err{18450}{2167}{2103}$ & $\err{2.06}{0.16}{0.18}$ & $\err{0.06}{0.04}{0.03}$ & $\err{2.37}{0.86}{0.86}$ & $\err{0.03}{0.38}{0.03}$ & $\err{0.17}{0.80}{0.17}$ & $\err{-3.44}{1.09}{1.23}$ & 1.7 \\
J0808+1728 & $\err{0.0862}{0.0170}{0.0122}$ & $\err{14250}{2088}{2162}$ & $\err{2.40}{0.35}{0.84}$ & $\err{0.14}{0.06}{0.10}$ & $\err{0.72}{0.99}{0.61}$ & $\err{0.18}{0.72}{0.18}$ & $\err{0.14}{1.60}{0.14}$ & $\err{-1.76}{0.88}{2.05}$ & 1.1 \\
J2136+0414 & $\err{0.0916}{0.0051}{0.0040}$ & $\err{10980}{1804}{1480}$ & $\err{2.45}{0.16}{0.15}$ & $\err{0.28}{0.02}{0.05}$ & $\err{4.31}{2.29}{1.97}$ & $\err{3.95}{0.47}{0.94}$ & $\err{0.05}{0.43}{0.05}$ & $\err{-4.83}{1.65}{1.92}$ & 12 \\
J0210-0124 & $\err{0.1500}{0.0000}{0.0008}$ & $\err{17120}{675}{849}$ & $\err{0.11}{0.30}{0.11}$ & $\err{0.41}{0.02}{0.01}$ & $\err{0.01}{0.05}{0.01}$ & $\err{0.00}{0.01}{0.00}$ & $\err{0.01}{0.05}{0.01}$ & $\err{-2.13}{0.17}{0.17}$ & 850 \\
%J2104-0035 & &&&&&&&\\
J0159-0622 & $\err{0.0882}{0.0019}{0.0020}$ & $\err{12720}{1027}{927}$ & $\err{1.83}{0.20}{0.29}$ & $\err{0.02}{0.01}{0.01}$ & $\err{0.49}{0.54}{0.46}$ & $\err{0.70}{0.17}{0.17}$ & $\err{3.45}{0.36}{0.36}$ & $\err{-5.43}{1.23}{0.57}$ & 27 \\
J0134-0038 & $\err{0.0815}{0.0057}{0.0057}$ & $\err{10100}{1107}{692}$ & $\err{0.09}{0.50}{0.09}$ & $\err{0.16}{0.04}{0.04}$ & $\err{0.01}{0.86}{0.01}$ & $\err{0.07}{0.33}{0.07}$ & $\err{1.14}{1.47}{0.87}$ & $\err{-4.15}{1.90}{1.90}$ & 4.2 \\
J0811+4730 & $\err{0.0932}{0.0078}{0.0078}$ & $\err{17590}{1541}{1590}$ & $\err{1.75}{0.26}{0.49}$ & $\err{0.00}{0.01}{0.00}$ & $\err{0.23}{0.63}{0.23}$ & $\err{0.00}{0.51}{0.00}$ & $\err{3.55}{1.28}{1.29}$ & $\err{-6.00}{1.15}{0.00}$ & 13 \\
J0845+0131 & $\err{0.1001}{0.0032}{0.0032}$ & $\err{10010}{816}{392}$ & $\err{0.43}{0.50}{0.39}$ & $\err{0.37}{0.02}{0.02}$ & $\err{1.45}{1.20}{0.95}$ & $\err{0.01}{0.31}{0.01}$ & $\err{0.82}{0.74}{0.52}$ & $\err{-1.71}{1.61}{1.98}$ & 110 \\
J2314+0154 & $\err{0.0783}{0.0057}{0.0057}$ & $\err{12830}{2201}{2278}$ & $\err{1.87}{0.70}{0.83}$ & $\err{0.27}{0.04}{0.05}$ & $\err{5.21}{1.96}{2.52}$ & $\err{3.90}{0.69}{1.21}$ & $\err{4.44}{0.56}{1.75}$ & $\err{-5.31}{1.34}{0.69}$ & 1.5 \\
J0007+0226 & $\err{0.0995}{0.0083}{0.0063}$ & $\err{17950}{1784}{1773}$ & $\err{1.97}{0.14}{0.15}$ & $\err{0.04}{0.04}{0.04}$ & $\err{0.03}{1.18}{0.03}$ & $\err{1.82}{1.03}{0.78}$ & $\err{2.17}{0.76}{0.76}$ & $\err{-2.53}{0.71}{1.94}$ & 1.9 \\
J0014-0043 & $\err{0.0812}{0.0071}{0.0043}$ & $\err{10420}{1396}{1178}$ & $\err{2.00}{0.31}{0.66}$ & $\err{0.17}{0.03}{0.08}$ & $\err{0.26}{0.63}{0.26}$ & $\err{0.20}{0.23}{0.18}$ & $\err{1.19}{0.67}{0.56}$ & $\err{-4.88}{1.34}{2.50}$ & 2.1 \\
J0226-5017 & $\err{0.1203}{0.0078}{0.0098}$ & $\err{16820}{2293}{2247}$ & $\err{1.11}{0.60}{0.48}$ & $\err{0.00}{0.03}{0.00}$ & $\err{2.79}{0.30}{0.32}$ & $\err{0.01}{0.29}{0.01}$ & $\err{0.34}{1.72}{0.34}$ & $\err{-4.49}{1.01}{0.89}$ & 5.0 \\
J0228-0210 & $\err{0.1360}{0.0140}{0.0610}$ & $\err{17610}{2976}{3676}$ & $\err{1.56}{0.56}{0.77}$ & $\err{0.21}{0.24}{0.12}$ & $\err{0.07}{1.31}{0.07}$ & $\err{0.02}{0.69}{0.02}$ & $\err{1.78}{1.77}{1.20}$ & $\err{-1.37}{0.74}{3.53}$ & 6.1 \\
J0248-0817 & $\err{0.0861}{0.0015}{0.0012}$ & $\err{14330}{635}{660}$ & $\err{1.87}{0.06}{0.07}$ & $\err{0.26}{0.01}{0.02}$ & $\err{0.14}{0.43}{0.14}$ & $\err{0.02}{0.12}{0.02}$ & $\err{0.00}{0.12}{0.00}$ & $\err{-4.95}{1.40}{1.35}$ & 39 \\
SBS0335E & $\err{0.0648}{0.0024}{0.0026}$ & $\err{14450}{877}{730}$ & $\err{2.50}{0.06}{0.05}$ & $\err{0.01}{0.01}{0.01}$ & $\err{0.13}{0.03}{0.02}$ & $\err{0.00}{0.01}{0.00}$ & $\err{4.98}{0.02}{0.07}$ & $\err{-3.78}{0.93}{0.95}$ & 120 \\
J0833+2508 & $\err{0.1010}{0.0123}{0.0113}$ & $\err{15170}{4482}{3407}$ & $\err{1.00}{0.70}{0.53}$ & $\err{0.02}{0.08}{0.02}$ & $\err{0.08}{0.50}{0.08}$ & $\err{0.02}{0.26}{0.02}$ & $\err{3.84}{1.61}{1.65}$ & $\err{-1.58}{1.33}{1.78}$ & 430 \\
Mrk 996 & $\err{0.1325}{0.0111}{0.0141}$ & $\err{10200}{2615}{2047}$ & $\err{2.84}{0.13}{0.13}$ & $\err{0.50}{0.00}{0.02}$ & $\err{0.04}{0.12}{0.04}$ & $\err{0.27}{0.12}{0.11}$ & $\err{4.50}{0.43}{0.64}$ & $\err{-1.35}{0.96}{1.34}$ & 48 \\
J0335-0038 & $\err{0.0829}{0.0080}{0.0055}$ & $\err{12340}{2031}{1880}$ & $\err{2.14}{0.31}{0.62}$ & $\err{0.28}{0.03}{0.04}$ & $\err{1.24}{0.83}{0.73}$ & $\err{0.01}{0.52}{0.01}$ & $\err{3.36}{1.54}{1.64}$ & $\err{-5.81}{1.54}{0.19}$ & 4.4 \\
J0301+0114 & $\err{0.0905}{0.0087}{0.0066}$ & $\err{15790}{1649}{1666}$ & $\err{1.69}{0.62}{0.81}$ & $\err{0.24}{0.03}{0.05}$ & $\err{1.84}{1.23}{0.97}$ & $\err{0.06}{0.92}{0.06}$ & $\err{1.22}{1.55}{1.14}$ & $\err{-5.11}{1.55}{0.89}$ & 0.48 \\
J0313+0006 & $\err{0.0693}{0.0052}{0.0032}$ & $\err{19850}{2094}{2008}$ & $\err{1.81}{0.35}{0.74}$ & $\err{0.30}{0.10}{0.11}$ & $\err{2.91}{1.71}{1.32}$ & $\err{0.04}{0.15}{0.04}$ & $\err{0.27}{0.65}{0.27}$ & $\err{-2.88}{1.40}{1.18}$ & 16
 \\
J0825+1846 & $\err{0.0727}{0.0031}{0.0028}$ & $\err{10440}{831}{440}$ & $\err{2.36}{0.15}{0.15}$ & $\err{0.11}{0.02}{0.02}$ & $\err{9.93}{0.07}{0.52}$ & $\err{1.63}{0.36}{0.34}$ & $\err{4.98}{0.02}{0.21}$ & $\err{-1.09}{0.09}{2.59}$ & 81 \\
J0815+2156 & $\err{0.0885}{0.0055}{0.0047}$ & $\err{17920}{1924}{1836}$ & $\err{1.85}{0.20}{0.24}$ & $\err{0.10}{0.03}{0.03}$ & $\err{2.27}{1.26}{1.10}$ & $\err{0.70}{0.57}{0.49}$ & $\err{1.74}{1.34}{1.06}$ & $\err{-4.39}{1.36}{1.43}$ & 2.5 \\
I Zw 18 NW & $\err{0.0711}{0.0033}{0.0024}$ & $\err{19950}{1295}{1682}$ & $\err{0.17}{0.36}{0.17}$ & $\err{0.05}{0.02}{0.03}$ & $\err{0.01}{0.14}{0.01}$ & $\err{0.17}{0.04}{0.04}$ & $\err{0.13}{0.29}{0.13}$ & $\err{-3.18}{0.61}{1.57}$ & 4.9 \\
\enddata
\end{deluxetable*}

\vspace{5mm}

\bibliography{reference}{}
\bibliographystyle{aasjournal}

\end{document}